\documentclass[aps,reprint,nofootinbib]{revtex4-1}
\usepackage{graphicx}
\usepackage{amsmath}
\usepackage{amsfonts}
\usepackage{amssymb}
\usepackage{bm}
\newcommand{\myarray}[1]{\begin{equation}\begin{split}#1\end{split}\end{equation}}

\begin{document}
\title{The Simplest Form of the Lorentz Transformations}
\date{\today}
\author{C. Baumgarten}
\affiliation{Switzerland}
\email{christian.baumgarten@gmx.net}

\def\begeq{\begin{equation}}
\def\endeq{\end{equation}}
\def\bquo{\begin{quotation}}
\def\equo{\end{quotation}}
\def\begary{\begeq\begin{array}}
\def\endary{\end{array}\endeq}
\def\bmtx{\left(\begin{array}}
\def\emtx{\end{array}\right)}
\def\eps{\varepsilon}
\def\d{\partial}
\def\y{\gamma}
\def\e{\eta}
\def\w{\omega}
\def\W{\Omega}
\def\s{\sigma}
\def\ket#1{\left|\,#1\,\right>}
\def\bra#1{\left<\,#1\,\right|}
\def\bracket#1#2{\left<\,#1\,\vert\,#2\,\right>}
\def\erw#1{\left<\,#1\,\right>}

\def\Exp#1{\exp\left(#1\right)}
\def\Log#1{\ln\left(#1\right)}
\def\Sinh#1{\sinh\left(#1\right)}
\def\Sin#1{\sin\left(#1\right)}
\def\Tanh#1{\tanh\left(#1\right)}
\def\Tan#1{\tan\left(#1\right)}
\def\Cos#1{\cos\left(#1\right)}
\def\Cosh#1{\cosh\left(#1\right)}

\begin{abstract}
We report the simplest possible form to compute rotations
around arbitrary axis and boosts in arbitrary directions 
for 4-vectors (space-time points, energy-momentum) and bi-vectors (electric
and magnetic field vectors) by symplectic similarity transformations.
The Lorentz transformations are based exclusively on real $4\times 4$-matrices 
and require neither complex numbers nor special implementations of abstract 
entities like quaternions or Clifford numbers. No raising or lowering of indices
is necessary. It is explained how the Lorentz transformations can be derived
from the most simple second order Hamiltonian of general significance.
Since this approach exclusively uses the {\it real} Clifford algebra $Cl(3,1)$,
all calculations are based on real $4\times 4$ matrix algebra.
\end{abstract}

\maketitle

\section{Introduction}

\begin{quotation}
A great many derivations of the Lorentz transformation have already
been given, and the subject, because of its pedagogical importance,
still receives continues attention [...]. Most of the analyses, following
the original one by Einstein, rely on the invariance of the speed of light
$c$ as a central hypothesis. That such an hypothesis, firmly based on 
experimental grounds, has had a crucial historical role cannot be denied.
The chronological building of order of a physical theory, however,
rarely coincides with its logical structure.\\ 
\mbox{}\hfill{\it -- J.M. Levy-Leblond~\cite{LevyLeblond}}
\end{quotation}

P.A.M. Dirac, the discoverer of relativistic quantum theory, wrote
that the ``real importance of Einstein's work was that he introduced
Lorentz transformations as something fundamental in physics''~\cite{Dirac80}.
But, as we shall argue, it is Dirac's theory and not Einstein's, which 
uncovers that the Lorentz transformations are indeed as fundamental as
Hamiltonian functions are fundamental, first of all in a mathematical,
but consequently also in a physical sense.

The Lorentz transformations (rotations and boosts) can be expressed using
different (though related) formulations. The respective form 
mainly depends on the type of vectorial system used to represent space and time
coordinates~\footnote{For the history of the different representations
see~\cite{Crowe,Chappell}.}. The most commonly promoted formulation of the 
Lorentz covariance are the vector and it's generalization, the tensor formalism.
As we shall demonstrate, these are neither algorithmically nor conceptually
the simplest variant.

We shall demonstrate here that the simplest possible form of the Lorentz 
transformations (LTs) is a direct consequence of the use of Hamiltonian 
methods. It is the irreducible remainder after a visit in Ockham's barber shop.
Our approach follows the work of Kim and Noz~\cite{Kim1981} and is 
closely related to (and inspired by) Dirac's equation, Hestenes' and Sobczyk' 
space-time algebra (STA)~\cite{STA,GA} and other 
Clifford algebraic approaches like the ones of Baylis~\cite{Baylis} or 
Salingaros~\cite{Salingaros}. 
However, our presentation differs from most others insofar as we derive a 
representation of the Lorentz transformations (LTs) directly from
Hamiltonian methods by the use of $4\times 4$ Dirac matrices over the 
{\it reals}. This matrix form is physically significant as the LTs are 
shown to be isomorphic to general linear canonical transformations of a 
acting on two coupled canonical pairs $(q_1,p_1,q_2,p_2)$. This kind of 
transformation is also called symplectic similarity transformation~\cite{Kim1981}.

In this representation physical observables like momentum and energy are not
regarded as self-sufficient ``fundamental'' quantities. Instead they are
related to (linear combinations of) second moments of phase-space
distributions (see Ref.~\cite{GLT}). In two previous publications we
explained that and how this reinterpretation of the LTs leads to 
a reinterpretation of quantum electrodynamics as a science of statistical 
moments in spinorial phase space~\cite{qed_paper,osc_paper}. The main
advantage of this approach is that all central quantities that determine the
motion of a charged particle in an electromagnetic field, including their
precise relations, can be derived from a single conservation law, namely 
in the form of the classical Hamiltonian function of two coupled harmonic
oscillators. The resulting form of the LTs is extraordinarily simple and 
straightforward.  
\bquo
I apologize, but theoretical physics is defined as a sequence of courses, 
each of which discusses the harmonic oscillator.\\
\mbox{}\hfill{\it -- Sidney Coleman~\cite{ColemanQFT}}
\equo
In order to motivate our approach we describe the conventional vector formalism
(CVF) and contrast it with the suggested formalism of symplectic similarity 
transformations in some detail.

\section{Space described by Vectors}
\label{sec_CVS}

A position or direction in space is most commonly represented by a vector.
As well-known, in CVF a ``vector'' is represented by a $3\times 1$-matrix
\begeq
{\bf x}=\bmtx{c}x\\y\\z\emtx\,,
\label{eq_vec1}
\endeq
or ${\bf x}=(x,y,z)^T$ with the superscipt ``T'' for matrix transposition.

If we construct unit vectors in each direction, then we may write:
\begeq
{\bf x}=x\,{\bf e}_x+y\,{\bf e}_y+z\,{\bf e}_z
\label{eq_vec2}
\endeq
where 
\begary{rclrclrcl}
{\bf e}_x&=&\bmtx{c}1\\0\\0\emtx\,,&
{\bf e}_y&=&\bmtx{c}0\\1\\0\emtx\,,&
{\bf e}_z&=&\bmtx{c}0\\0\\1\emtx\\
\endary
The scalar product (dot product) of two vectors can be implemented as a 
product of a transposed $3\times 1$-matrix times a $3\times 1$-matrix
\begeq
{\bf x}_1\cdot{\bf x}_2={\bf x}_1^T\,{\bf x}_2=x_1\,x_2+y_1\,y_2+z_1\,z_2\,.
\endeq
Unfortunately, this form to represent a vector has the undesired feature that
the scalar multiplication changes the algebraic dimension and yields - as the
name suggests - a scalar. Strange enough, there is a second type of 
vector multiplication, the so-called ``vector'' or ``cross'' 
product, which requires an extra symbol, namely the cross, and has its own 
definition:
\begary{rcl}
{\bf x}_1\times {\bf x}_2&=&(y_1\,z_2-y_2\,z_1)\,{\bf e}_x\\
                         &+&(z_1\,x_2-z_2\,x_1)\,{\bf e}_y\\
                         &+&(x_1\,y_2-x_2\,y_1)\,{\bf e}_z\\
\label{eq_crossprod}
\endary
At first sight the cross product is an speciality of $3$-dimensional 
space and has no generalization to arbitrary dimensions and no obvious place 
within a generalized vector- and matrix-algebra. However, the cross product is 
physically and geometrically indispensable. It represents real and measurable 
properties of $3$-dimensional physical space, namely the {\it handedness} of magnetic 
and gyroscopic forces. The need to define two different products indicates, that 
an unstructured ``list'' of coordinates does not adequately represent the 
structural properties of $3$-dimensional ``physical'' space.

\subsection{Rotations}
\label{sec_vecrot}

\begin{figure}
\includegraphics[width=8cm]{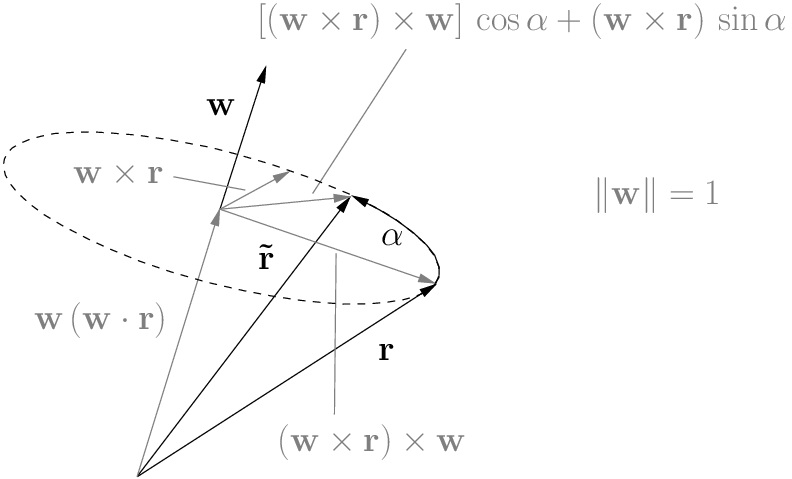}
\caption{
Rotation of an arbitrary vector ${\bf r}$ around arbitrary unit vector ${\bf w}$ with
angle $\alpha$.
\label{fig_rot}
}
\end{figure}

Let us consider the rotation of a vector ${\bf r}$ by an angle
$\alpha$ about an arbitrary direction indicated by the unit vector
${\bf w}$. The derivation of an appropriate formula requires the computation 
of the vector-components parallel and perpendicular to ${\bf w}$ and it is helpful
to use a drawing that clarifies the situation (see Fig.~\ref{fig_rot}). 
Besides the $\sin()$- and $\cos()$-function mainly vector addition 
and the computation of scalar and cross-products are needed in order to
decompose the vector into the component parallel and perpendicular to ${\bf
 w}$, respectively.
\begary{rcl}
{\bf r}&=&{\bf r}_\parallel+{\bf r}_\perp\\
{\bf r}_\parallel&=&({\bf w}\cdot{\bf r})\,{\bf w}\\
{\bf r}_\perp&=&({\bf w}\times{\bf r})\times{\bf w}\\
             &=&{\bf r}-{\bf r}_\parallel={\bf r}-({\bf w}\cdot{\bf r})\,{\bf w}\\
{\bf\tilde r}&=&{\bf r}_\parallel+{\bf r}_\perp\,\cos{\alpha}+({\bf w}\times{\bf r})\sin{\alpha}\,.
\endary
From this we can derive the most simple formula of CVF, the formula of Rodriguez: 
\begeq
{\bf\tilde r}={\bf r}\,\cos{\alpha}+({\bf w}\times{\bf r})\sin{\alpha}+{\bf
  w}\,({\bf w}\cdot{\bf r})\,(1-\cos{\alpha})\,.
\label{eq_rodrigues}
\endeq
For the description of a supposedly fundamental operation like rotation
in space, this formula is surprizingly complicate.

An alternative approach is the use of matrices to describe rotations. Since positions
are represented in CVF by $3\times 1$ matrices, the rotation matrices
${\bf Q}_x$, ${\bf Q}_y$ and ${\bf Q}_z$ are orthogonal matrices of dimension
$3\times 3$
\begeq
{\bf\tilde x}={\bf Q}_k(\alpha)\,{\bf x}\,.
\endeq
where $k$ indicates a rotation axis.
These rotation matrices are
\begary{rcl}
{\bf Q}_x&=&\bmtx{ccc}
1&0&0\\
0&\cos{\alpha_1}&-\sin{\alpha_1}\\
0&\sin{\alpha_1}&\cos{\alpha_1}\\
\emtx\\
{\bf Q}_y&=&\bmtx{ccc}
\cos{\alpha_2}&0&\sin{\alpha_2}\\
0&1&0\\
-\sin{\alpha_2}&0&\cos{\alpha_2}\\
\emtx\\
{\bf Q}_z&=&\bmtx{ccc}
\cos{\alpha_3}&-\sin{\alpha_3}&0\\
\sin{\alpha_3}&\cos{\alpha_3}&0\\
0&0&1\\
\emtx\\
\endary
${\bf Q}_x$, ${\bf Q}_y$ and ${\bf Q}_z$ represent rotations around the coordinate axis
${\bf e}_x$, ${\bf e}_y$ and ${\bf e}_z$. These matrices can be obtained from the matrix 
exponential of ``infinitesimal'' rotations ${\bf R}_k$, which are simply the derivatives
of the ${\bf Q}_k$:
\begeq
{\bf R}_k=\left.{d\over d\alpha_k}{\bf Q}_k(\alpha_k)\right\vert_{\alpha_k=0}\,.
\endeq
such that 
\begeq
{\bf R}_x=\bmtx{ccc}
0&  0 & 0\\
0&  0 &-1\\
0&  1 & 0\\
\emtx\,.
\endeq
An infinitesimal rotation is then given by:
\begeq
{\bf\tilde x}=({\bf 1}+{\bf R}_k\,d\alpha_k)\,{\bf x}\,.
\endeq
The description of a general rotation of the vector ${\bf x}$ around an arbitrary axis 
$\vec\w$ (with $\vert\vec\w\vert=1$) can be done by a single matrix multiplication with a matrix 
${\bf Q}$ which can be computed as the matrix exponential of the corresponding infinitesimal
transformation:
\begeq
{\bf Q}=\exp{((\w_x\,{\bf R}_x+\w_y\,{\bf R}_y+\w_z\,{\bf R}_z)\,\alpha)}
\endeq 
It is explicitely given by
\begeq
{\bf \tilde x}=
\bmtx{c}\tilde x\\\tilde y\\\tilde z\emtx=
\bmtx{ccc}
Q_{xx}&Q_{xy}&Q_{xz}\\
Q_{yx}&Q_{yy}&Q_{yz}\\
Q_{zx}&Q_{zy}&Q_{zz}\\
\emtx\,
\bmtx{c}x\\y\\z\emtx
\label{eq_LT_CVF}
\endeq
where
\begary{rcl}
Q_{xx}&=& c+(1-c)\,\w_x^2\\
Q_{xy}&=&(1-c)\,\w_x\,\w_y-s\,\w_z\\
Q_{xz}&=&(1-c)\,\w_x\,\w_z+s\,\w_y\\
Q_{yx}&=&(1-c)\,\w_x\,\w_y+s\,\w_z\\
Q_{yy}&=& c+(1-c)\,\w_y^2\\
Q_{yz}&=&(1-c)\,\w_y\,\w_z-s\,\w_x\\
Q_{zx}&=&(1-c)\,\w_x\,\w_z-s\,\w_y\\
Q_{zy}&=&(1-c)\,\w_y\,\w_z+s\,\w_x\\
Q_{zz}&=&c+(1-c)\,\w_z^2\\
\endary
and $c=\cos{(\alpha)}$ and $s=\sin{(\alpha)}$. This way to describe rotations can, in principle,
be extended to arbitrary dimensions, which means that it has {\it no intrinsic connection to 
the dimensionality of space}.

Surprisingly enough, the conventional rotation matrices ${\bf Q}_k$ are not directly used to 
describe the motion of rigid bodies in $3$-dimensional space. Instead, most textbooks 
recommend the use of Euler angles. The Euler angles are a powerful tool, but again are 
not simple or intuitive: Greiner for instance explains these angles with three 
figures~\cite{Greiner}.
Even though the human mind is trained to grasp 3-dimensional situations, when it comes
to real calculations, 3-dimensional space is remarkably tedious. This becomes even worse
when Lorentz boosts and electromagnetic fields are considered as we shall see in Sec.~\ref{sec_boost}.

In the Hamiltonian Clifford algebra (HCA) suggested here, unit ``vectors'' are
represented by real $4\times 4$-matrices $\y_k$ and the rotation
of an arbitrary vector ${\bf x}=x\,\y_1+y\,\y_2+z\,\y_3$ around an arbitrary 
direction is generated by this same ``direction''
\begeq
{\bf w}=\w_x\,\y_7+\w_y\,\y_8+\w_z\,\y_9 
\label{eq_gam789}
\endeq
applied to ${\bf x}$ in the form of a similarity transformation
\begeq
{\bf\tilde x}={\bf R}\,{\bf x}\,{\bf R}^{-1}
\label{eq_LT}
\endeq
One should not be confused by the wording of ``vector'' and ``matrix''.
In the CVF, a ``vector'' is formally a column ``matrix''. In the approach 
suggested here, a ``vector'' has the algebraic form of a matrix, not of a 
column matrix, but of a real $4\times 4$ matrix. This matrix may contain more 
information than that of a single column-``vector'' and, as we shall show, 
it can be used to represent the structure of space-time and (quantum-)
electrodynamics. Mathematically $4\times 4$ matrices can be used to 
represent specific Clifford algebras.
But in case of matrices of dimension $2^N\times 2^N$, the reverse is true as
well: Such matrices can always be represented in terms of Clifford algebras.
This means that any real $2^N\times 2^N$-matrix can be expressed as a weighted
sum of rather simple elementary matrices.
We shall explain this in more detail in Sec.~\ref{sec_matrixrep}.
 
Using a Clifford algebraic matrix decomposition, the transformation matrix 
${\bf R}$ is again a matrix exponential of the generator ${\bf w}$
\begeq
{\bf R}=\exp{(-{\bf w}\,\alpha/2)}\,.
\label{eq_expo}
\endeq
The ``vector'' ${\bf w}$, which represents the direction of rotation, 
has the same form as the ``vector'' ${\bf x}$, namely that of a 
$4\times 4$ Hamiltonian matrix. The unit matrices $\y_7$, $\y_8$ and $\y_9$
are simply products of {\it two} real Dirac matrices and are therefore called
``bi-vectors''. They are defined by
\begary{rcl}
\y_7&=&\y_2\,\y_3\\
\y_8&=&\y_3\,\y_1\\
\y_9&=&\y_1\,\y_2\,.
\label{eq_gam789p}
\endary
the form and meaning of which will be explained later.

All generators of rotations (like ${\bf w}$) square to $-{\bf 1}$ 
(i.e. are representations of the unit imaginary $i$), such that Eq.~\ref{eq_expo} 
yields Eulers formula:
\begeq
{\bf R}=\cos{(\alpha/2)}\,{\bf 1}-\sin{(\alpha/2)}\,{\bf w}\,.
\endeq
The inverse transformation is given by the negative argument
${\bf R}^{-1}(\alpha)={\bf R}(-\alpha)$:
\begeq
{\bf R}^{-1}=\cos{(\alpha/2)}\,{\bf 1}+\sin{(\alpha/2)}\,{\bf w}\,.
\endeq
The explicit form of the matrix is, in the chosen representation, given by:
\begary{rcl}
{\bf R}(\alpha)&=&{\bf 1}\,\cos{(\alpha/2)}-{\bf w}\,\sin{(\alpha/2)}\\
       &=&\bmtx{cccc}
c & -\w_y\,s & \w_z\,s&-\w_x\,s\\
\w_y\,s & c &\w_x\,s&\w_z\,s\\
-\w_z\,s & -\w_x\,s& c &\w_y\,s\\
\w_x\,s & -\w_z\,s&-\w_y\,s &c\\
\emtx
\endary
where $c=\cos{(\alpha/2)}$, $s=\sin{(\alpha/2)}$.

Rotations, when expressed by a similarity transformation
(Eq.~\ref{eq_LT}), require {\it two} instead of one matrix 
multiplication(s) as in Eq.~\ref{eq_LT_CVF}. One might therefore 
doubts that this symplectic method is really ``simpler''.
But firstly Eq.~\ref{eq_LT} can simultaneously be used to rotate not
only the vector ${\bf x}$, but two bi-vectors as well, i.e.
three different ``vectors''. Secondly, the exact same form
of matrix multiplication can be used to compute Lorentz boosts as
well, as we shall demonstrate next.

Thirdly the use of the Hamiltonian Clifford algebras allows to relate
geometrical to {\it dynamical} concepts. And these concepts can be 
derived logically within linear Hamiltonian theory with a minimal 
number of assumptions. And finally it exemplifies a considerable 
number of concepts used in modern mathematical physics
in one go, including group and representation theory, Clifford algebras,
symplectic motion, canonical transformations up to the Lorentz
covariance of the Dirac equation. It therefore has unique educational 
value.

\subsection{Lorentz Boost of $4$-vectors}
\label{sec_boost}

Jacksons ``Electrodynamics'' presents the following formula, with the restriction that 
the boost must be along $z$~\cite{Jackson}:
\begary{rcl}
z'&=&{z-v\,t\over\sqrt{1-\frac{v^2}{c^2}}}\\
t'&=&{t-\frac{v}{c^2}\,z\over\sqrt{1-\frac{v^2}{c^2}}}\\
x'&=&x\\
y'&=&y\\
\label{eq_LTvector}
\endary
and, for the general case:
\begary{rcl}
{\bf x'}_{\parallel}&=&{1\over\sqrt{1-\frac{v^2}{c^2}}}\,({\bf x}_{\parallel}-{\bf v}\,t)\\
t'&=&{1\over\sqrt{1-\frac{v^2}{c^2}}}\,(t-\frac{{\bf v}\cdot{\bf x}}{c^2})\\
{\bf x'}_\perp&=&{\bf x}_\perp\\
\endary
Again it is required to split vectors into the parallel and perpendicular 
components. The conventional matrix formalism, as an extension of CVF (ECVF), requires now the
use of $4\times 4$-matrices, the ``4-vector'' ${\bf x}=(t,x,y,z)^T$ has four components.

The infinitesimal generator of a boost in direction $\vec\w$ ($\vec\w^2=1$) is then a symmetric matrix:
\begary{rcl}
{\bf B}&=&\bmtx{cccc}
0 & \w_x & \w_y & \w_z\\
\w_x & 0 & 0 & 0\\
\w_y & 0 & 0 & 0\\
\w_z & 0 & 0 & 0\\
\emtx\\
\endary
The matrix exponential required for a boost with finite velocity
\begary{rcl}
{\bf L}&=&\exp{({\bf B}\,\tau)}\\
{\bf x}'&=&{\bf L}\,{\bf x}
\endary
with $\cosh{\tau}=\y$ and $\beta\y=\sinh{\tau}$ is given by
\begary{rcl}
{\bf L}&=&\bmtx{cccc}
L_{tt} & L_{tx} &L_{ty} &L_{tz} \\
L_{tx} & L_{xx} &L_{xy} &L_{xz} \\
L_{ty} & L_{xy} &L_{yy} &L_{yz} \\
L_{tz} & L_{xz} &L_{yz} &L_{zz} \\
\emtx\,.
\endary
The matrix elements are:
\begary{rcl}
L_{tt}&=&\y\\
L_{xx}&=& 1 + (\y-1)\,\w_x^2\\
L_{xx}&=& 1 + (\y-1)\,\w_y^2\\
L_{xx}&=& 1 + (\y-1)\,\w_z^2\\
\endary
and
\begary{rclp{5mm}rcl}
L_{tx}&=&  -\y\,\beta\,\w_x  && L_{xy}&=&(\y-1)\,\w_x\,\w_y\\
L_{ty}&=&-\y\,\beta\,\w_y    && L_{xz}&=& (\y-1)\,\w_x\,\w_z \\
L_{tz}&=&  -\y\,\beta\,\w_z  && L_{yz}&=&(\y-1)\,\w_y\,\w_z\\ 
\endary
where $\w_x^2+\w_y^2+\w_z^2=1$.

The conventional presentation of special relativity gives no 
{\it logical} argument why space-time should have a Minkowski type geometry
and no reason why space-time {\it should} have $3+1$ dimensions: At first 
sight it seems straightforward to extend this formalism to any number of 
spatial and temporal dimensions by extending the size of the rotation and 
boost matrices. The Hamiltonian Clifford algebra, suggested here, is in this
respect considerably more restrictive. This provides a degree of explanatory 
power that the CVF can not provide~\cite{qed_paper}.

The boost of an arbitrary 4-vector is, yet again, performed by a boost matrix 
${\bf B}$ in the form of a symplectic similarity transformation ${\bf x}\to{\bf x}'$:
\begeq
{\bf x'}={\bf B}\,{\bf x}\,{\bf B}^{-1}
\endeq
where the boost matrix ${\bf B}$ is, yet again, given by a matrix exponential
\begeq
{\bf B}=\exp{(-{\bf\eps}\,\tau/2)}
\endeq
in which the infinitesimal generator $\eps$ has the same structure as ${\bf x}$, 
namely that it is a $4\times 4$ Hamiltonian matrix. Generators of boosts $\eps$ 
square to ${\bf 1}$, such that the matrix exponential yields
\begeq
{\bf B}=\cosh{(\tau/2)}\,{\bf 1}-\sinh{(\tau/2)}\,\eps\,.
\endeq
Again the inverse matrix is given by the negative argument 
${\bf B}^{-1}(\tau)={\bf B}(-\tau)$. The sign of the squared
generator is the (only) significant {\it formal} difference 
between rotations and boosts.

The matrix $\eps$ is again essentially a direction ``vector''
\begeq
\eps=\eps_x\,\y_4+\eps_y\,\y_5+\eps_z\,\y_6
\label{eq_boost}
\endeq
where the unit matrices (yet again ``bi-vectors'') are 
\begary{rcl}
\y_4&=&\y_0\,\y_1\\
\y_5&=&\y_0\,\y_2\\
\y_6&=&\y_0\,\y_3\,.
\endary
If we use a normalization $\vert\vec\eps\vert=1$, then
the matrix $\eps$ squares, in contrast to the generators of rotations, 
to the positive unit matrix $\eps^2=+{\bf 1}$, which characterizes 
these matrices as generators of boosts. 
The matrix exponent is then explicitely given by:
\begary{rcl}
{\bf B}(\tau)&=&{\bf 1}\,\cosh{(\tau/2)}-\eps\,\sinh{(\tau/2)}\\
       &=&\bmtx{cccc}
c+\eps_x\,s & -\eps_z\,s & -\eps_y\,s & 0\\
-\eps_z\,s & c-\eps_x\,s & 0 & \eps_y\,s \\
-\eps_y\,s& 0 & c-\eps_x\,s & -\eps_z\,s \\
0 & \eps_y\,s & -\eps_z\,s & c+s\,\eps_x \\
\emtx
\endary
where $c=\cosh{(\tau/2)}$ and $s=\sinh{(\tau/2)}$. The parameter
$\tau$ is the ``rapidity''; in conventional notation with
$\beta=v/c$ and $\y=1/\sqrt{1-\beta^2}$ one has
\begary{rcl}
\cosh{(\tau)}&=&\y\\
\sinh{(\tau)}&=&\beta\,\y\\
\tanh{(\tau)}&=&\beta\\
\endary
In contrast to the usual tensor formalism which essentially has to be 
learned and memorized, the approach suggested here can be logically 
developed from little more than a single conservation law~\cite{qed_paper}.
To memorize it, it suffices to understand the (classical Hamiltonian) 
principles underlying this approach.  

\subsection{Lorentz Boost of Electromagnetic Fields}

So far our treatment concerned only the transformations of ``vector'' components.
Now we include electromagnetic fields. The corresponding formulas are, again 
assumed that the parallel and perpendicular components are computed 
beforehand~\cite{Jackson}:
\begary{rcl}
\gamma&=&{1\over\sqrt{1-\frac{{\bf v}^2}{c^2}}}\\
{\bf E'}_{\parallel}&=&{\bf E}_{\parallel}\\
{\bf B'}_{\parallel}&=&{\bf B}_{\parallel}\\
{\bf E'}_{\perp}&=&\y\,({\bf E}_{\perp}+\frac{\bf v}{c}\times{\bf B})\\
{\bf B'}_{\perp}&=&\y\,({\bf B}_{\perp}-\frac{\bf v}{c}\times{\bf E})\\
\endary
Once again we have a new set of formulas, significantly different from 
Eq.~\ref{eq_LTvector}.
Apparently there are different types of ``vectors'' within the CVF, but the
CVF provides no means to distinguish or label these vector types formally.
Even though it is well-known that magnetic field ``vectors'' are ``axial''
and electric field vectors are ``radial'' vectors, the CVF represents
them all by $3\times 1$-column matrices. Without context, one can not
possibly decide which type of transformation has to be applied.
The conventional approach then introduces a tensor formalism and claims that 
the ``vectors'' ${\bf E}$ and ${\bf B}$ 
of the electromagnetic field are indeed not ``vectors'', but components 
of a tensor and that the transformation of this tensor ${\bf F}$ requires
- in contrast to the transformation of vector type elements -
a double multiplication with the transformation matrix according to
${\bf F'}={\bf L}\,{\bf F}\,{\bf L}^T$ (see Jackson~\cite{Jackson}, chap. 11).

There is no doubt that the tensor formalism is {\it mathematically} 
correct, but this formalism does not provide a {\it reason} why physical 
space should be just so.
Hence, with respect to logic and aesthetics, the conventional approach 
remains a patchwork of remarkable unseemliness, especially with respect 
to the procedures of ``raising'' and ``lowering'' of indices.

In the Hamiltonian Clifford Algebra described here, a boost of 
4-vectors as well as electromagnetic fields, is represented by a 
matrix ${\bf B}$ in the {\it same form}, namely that of a symplectic 
similarity transformation ${\bf F}\to{\bf\tilde F}$:
\begeq
{\bf\tilde F}={\bf B}\,{\bf F}\,{\bf B}^{-1}
\endeq
where the boost matrix ${\bf B}$ and the generator $\eps$ have already 
been given above: the energy-momentum 4-vector is transformed with the
same transformation matrices as the electromagnetic fields.
As already mentioned, the (Hamiltonian) matrix ${\bf F}$ has the capacity
to represent exactly for the required number of independent parameters, 
namely ten, to represent a 4-vector and six field components, the latter
being naturally grouped into two sets of three components. 
I.e. $4\times 4$ real matrices simultaneously contain a ``vector'' (called 
4-vector in the ECVF) and two so-called ``bi-vectors'' (i.e. a ``tensor'') 
also given above in Eq.~\ref{eq_boost} and Eq.~\ref{eq_rot}.
The use of complex numbers is not required. 
The combination of a simultaneous boost and rotation ${\bf (BR)}$ is, due 
to the ``superposition principle'', obtained as the matrix exponential of 
the sum of the generators:
\begary{rcl}
{\bf (BR)}=\exp{(-(\eps+{\bf w})\,\phi/2)}
\endary
The composition of the generators is simple and can be derived in a
straightforward manner from the algebraic structure of the phase space of two 
canonical pairs: the symplectic Hamiltonian Clifford algebra 
$Cl(3,1)$, which is represented by a complete set of $4\times 4$-matrices and 
is just a real-valued variant of the Dirac algebra. 

\section{Matrix Representations}
\label{sec_matrixrep}

Let us motivate the use of matrices representing unit directions
starting from the conventional vector formalism (CVF). We take a new look at (Eq.~\ref{eq_vec2}), 
i.e. at two ``vectors'' and their product
\begary{rcl}
{\bf x}_1&=&x_1\,{\bf e}_x+y_1\,{\bf e}_y+z_1\,{\bf e}_z\\
{\bf x}_2&=&x_2\,{\bf e}_x+y_2\,{\bf e}_y+z_2\,{\bf e}_z\\
{\bf x}_1\cdot{\bf x}_2&=&x_1\,x_2\,{\bf e}_x^2+y_1\,y_2\,{\bf e}_y^2+z_1\,z_2\,{\bf e}_z^2\\
&+&x_1\,y_2\,{\bf e}_x\cdot{\bf e}_y+y_1\,x_2\,{\bf e}_y\cdot{\bf e}_x\\
&+&x_1\,z_2\,{\bf e}_x\cdot{\bf e}_z+z_1\,x_2\,{\bf e}_z\cdot{\bf e}_x\\
&+&y_1\,z_2\,{\bf e}_y\cdot{\bf e}_z+z_1\,y_2\,{\bf e}_z\cdot{\bf e}_y\\
\label{eq1}
\endary
In the conventional formalism, unit vectors ${\bf e}_i$ are commuting 
and pairwise orthogonal $3$-vectors, so that
\begeq
{\bf e}_i\cdot{\bf e}_j={\bf e}_j\cdot{\bf e}_i=\delta_{ij}\,,
\label{eq_cvf_scalarprod}
\endeq
with the Kronecker $\delta_{ij}$~\footnote{
The Kronecker delta is defined by:
$\delta_{ij}=0$ for $i\ne j$ and $\delta_{ij}=1$ for $i=j$.
} and hence Eq.~\ref{eq1} reduces to the scalar product
\begeq
{\bf x}_1\cdot{\bf x}_2=x_1\,x_2+y_1\,y_2+z_1\,z_2\,,
\endeq
since all mixed terms in Eq.~\ref{eq1} vanish.

However, if we look more closely on Eq.~\ref{eq1}, we note 
that the cross product is already there, if the unit 
elements ${\bf e}_i$ do not commute, but anti-commute~\footnote{
This idea goes essentially back to Sir W.R. Hamilton who
published his discovery of the so-called quaternions
already in 1844~\cite{Quaternions}.}.
That is, if for $i\ne j$ we assume that
\begeq
{\bf e}_i\cdot{\bf e}_j=-{\bf e}_j\cdot{\bf e}_i
\endeq
then Eq.~\ref{eq_cvf_scalarprod} can be replaced by
\begeq
2\,\delta_{ij}={\bf e}_i\,{\bf e}_j+{\bf e}_j\,{\bf e}_i\,.
\label{eq_acom}
\endeq
In this case, since Eq.~\ref{eq_acom} implies that ${\bf e}_i^2={\bf 1}$, one finds
\begary{rcl}
{\bf x}_1\,{\bf x}_2&=&(x_1\,x_2+y_1\,y_2+z_1\,z_2)\,{\bf 1}\\
&+&(y_1\,z_2-y_2\,z_1)\,{\bf e}_y\,{\bf e}_z\\
&+&(x_2\,z_1-x_1\,z_2)\,{\bf e}_z\,{\bf e}_x\\
&+&(x_1\,y_2-x_2\,y_1)\,{\bf e}_x\,{\bf e}_y\,,
\label{eq_new_scalarprod}
\endary
where the bold-face ${\bf 1}$ represents a unit matrix.
The resulting expression then is a combination of the scalar 
and the vector product. This becomes more obvious, if we 
identify the products
\begary{rcl}
{\bf b}_x&=&{\bf e}_y\,{\bf e}_z\\
{\bf b}_y&=&{\bf e}_z\,{\bf e}_x\\
{\bf b}_z&=&{\bf e}_x\,{\bf e}_y\,,
\label{eq_rotators}
\endary
with a new type of (unit-) vector, the already mentioned ``bi-vector'',
already known from Eq.~\ref{eq_gam789} and Eq.~\ref{eq_gam789p}.

We can then redefine the scalar product by using the
anti-commutator of ${\bf x}_1$ and ${\bf x}_2$ according to
\begeq
{\bf x}_1\,{\bf x}_2+{\bf x}_2\,{\bf x}_1=2\,({\bf x}_1\cdot{\bf x}_2)\,{\bf 1}\,,
\endeq
which is still a (unit) matrix. In order to obtain a scalar, we computes the
trace of the matrix and divides it by the number $n$ of diagonal elements:
\begeq
({\bf x}_1\cdot{\bf x}_2)_S\equiv\frac{1}{2\,n}\,\mathrm{Tr}({\bf x}_1\,{\bf x}_2+{\bf x}_2\,{\bf x}_1)
\endeq
In some sense this establishes two types of orthogonality, a strong version
in which two matrices simply anticommute and a weak version, in which the 
anticommutator does not vanish, but is traceless. We call this second product
the inner product:
\begeq
{\bf x}_1\cdot{\bf x}_2\equiv\frac{1}{2}\,({\bf x}_1\,{\bf x}_2+{\bf x}_2\,{\bf x}_1)
\endeq

Accordingly, the ``vector product'' or outer product is, in this matrix-representation, 
given by
\begeq
{\bf x}_1\wedge{\bf x}_2\equiv\frac{1}{2}\,({\bf x}_1\,{\bf x}_2-{\bf x}_2\,{\bf x}_1)
\endeq
The trace of the commutator of two matrices is always zero and it would therefore
be meaningless to define something like an ``outer scalar product''.
The product of two matrices always involves both products:
\begeq
{\bf x}_1\,{\bf x}_2={\bf x}_1\cdot{\bf x}_2+{\bf x}_1\wedge{\bf x}_2\,.
\endeq

Since the unit ``vectors'' ${\bf e}_i$, represented by matrices,
anti-commute and square to ${\bf 1}$, the elements of the bi-vector 
${\bf b}$ square to $-{\bf 1}$:
\begary{rcl}
{\bf b}_x^2&=&{\bf e}_y\,{\bf e}_z\,{\bf e}_y\,{\bf e}_z\\
&=&-{\bf e}_y\,({\bf e}_z\,{\bf e}_z)\,{\bf e}_y\\
&=&-{\bf e}_y\,{\bf e}_y\\
&=&-{\bf 1}\\
\endary
and (as can easily be shown) they mutually anti-commute, just as we presumed
for the vector-type elements ${\bf e}_i$~\footnote{
Note that the bi-vector ${\bf b}$ is a representation of the quaternion
elements ${\bf i}$, ${\bf j}$ and ${\bf k}$.
}. Hence, if one finds three (orthogonal and therefore mutually anti-commuting) 
direction matrices ${\bf e}_x$, ${\bf e}_y$ and ${\bf e}_z$, then there are
at least three more anti-commuting matrices ${\bf b}_x$, ${\bf b}_y$ and ${\bf b}_z$,
which square to the negative unit matrix.

To those who are unfamiliar with classical mechanics and the fundamental importance
of the cross product for the description of angular momentum, gyroscopic forces and
magnetic fields, the representation of a direction by a matrix might at first sight 
appear as a somewhat artificial mathematical construction. 
But if one considers the Hamiltonian origin of this approach in some more
detail, then it turns out to be the simplest and most natural representation
of space and, as we shall demonstrate in the following, it automatically generates
the Lorentz transformations and (from a generalized perspective) provides arguments 
for the inevitable geometry and dimensionality of real ``physical'' space-time.
The Clifford algebra $Cl(3,1)$ provides a conceptual understanding of physical
space as a dynamical structure that can, in this simple form, not be obtained otherwise.

As mentioned before, it is a major advantage of the representation of spatial 
unit directions by real square matrices that all sums and products of square matrices 
are again square matrices of the same dimension. It is therefore possible to compute 
arbitrary analytical functions of square matrices in the form of Taylor series, for 
instance the matrix exponential, which is the natural form in any type of linear
non-degenerate evolution in time.
While computation of the matrix exponential of arbitrary Hamiltonian matrices
is - in the general case - quite involved~\cite{exp_paper}, it signi\-ficantly
simplifies, if the argument squares to the (positive or negative) unit matrix
${\bf b}^2=\pm{\bf 1}$. The Taylor series can then be splitted into the even and 
odd partial series, such that with 
${\bf b}^2=s\,{\bf 1}$ (with the sign $s=\pm 1$) one obtains:
\begary{rcl}
\exp{({\bf b}\,\phi)}&=&\sum\limits_{k=0}^\infty\,{({\bf b}\,\phi)^k\over k!}\\
&=&\sum\limits_{k=0}^\infty\,{({\bf b}\,\phi)^{2k}\over (2\,k)!}+\sum\limits_{k=0}^\infty\,{({\bf b}\,\phi)^{2k+1}\over (2\,k+1)!}\\
&=&{\bf 1}\,\sum\limits_{k=0}^\infty\,{s^k\,\phi^{2k}\over (2\,k)!}+{\bf b}\,\sum\limits_{k=0}^\infty\,{s^k\,\phi^{2k+1}\over (2\,k+1)!}\\
\endary
such that with $s=-1$ one finds
\begeq
{\bf R}=\exp{({\bf b}\,\phi)}={\bf 1}\,\cos{(\phi)}+{\bf b}\,\sin{(\phi)}\,,
\label{eq_rotgen}
\endeq
If a matrix ${\bf b}$ squares to the positive unit matrix, i.e. if $s=1$, then
it follows that
\begeq
{\bf B}=\exp{({\bf b}\,\phi)}={\bf 1}\,\cosh{(\phi)}+{\bf b}\,\sinh{(\phi)}\,.
\label{eq_boostgen}
\endeq
Obviously we have $(\exp{({\bf b}\,\phi)})^{-1}=\exp{(-{\bf b}\,\phi)}$. Furthermore, the exponential
of this type of ``unit matrices'' ${\bf b}$ is a linear combination of the unit matrix ${\bf 1}$ and ${\bf b}$
such that the matrices ${\bf b}$ and $\exp{({\bf b}\,\phi)}$ commute with the same matrices.

Consider the transformation of a ``vector'' ${\bf x}=x\,{\bf e}_x+ y\,{\bf e}_y+ z\,{\bf e}_z$ according to
\begary{rcl}
{\bf\tilde x}&=&{\bf R}\,{\bf x}\,{\bf R}^{-1}\\
&=&x\,{\bf R}\,{\bf e}_x\,{\bf R}^{-1}+y\,{\bf R}\,{\bf e}_y\,{\bf R}^{-1}+z\,{\bf R}\,{\bf e}_z\,{\bf R}^{-1}\,,
\label{eq_trafo0}
\endary
If the transformation matrix ${\bf R}$ {\it commutes} with ${\bf e}_i$, then this component is unchanged.
But what happens, if it does not commute?

\subsection{Rotations as Similarity Transformations}
\label{sec_rot}

Let us explicitely calculate the result of the transformation (Eq.~\ref{eq_trafo0})
with a rotation matrix ${\bf R}=\exp{(-{\bf b}\,\phi/2)}$. We use the abbreviations
$c=\cos{(\phi/2)}$, $s=\sin{(\phi/2)}$, $C=\cos{(\phi)}$ and $S=\sin{(\phi)}$:
\begeq
{\bf\tilde x}=({\bf 1}\,c-{\bf b}\,s)\,(x\,{\bf e}_x+ y\,{\bf e}_y+ z\,{\bf e}_z)\,({\bf 1}\,c+{\bf b}\,s)
\label{eq_rot}
\endeq
where ${\bf b}={\bf b}_z={\bf e}_x\,{\bf e}_y$, which we evaluate component-wise:
\begary{rcl}
{\bf\tilde x}_x&=&({\bf 1}\,c-{\bf b}\,s)\,x\,{\bf e}_x\,({\bf 1}\,c+{\bf b}\,s)\\
&=&x\,\left({\bf e}_x\,c^2-{\bf b}\,{\bf e}_x\,{\bf b}\,s^2+({\bf e}_x\,{\bf b}-{\bf b}\,{\bf e}_x)\,c\,s)\right)\\
\endary
Now, the anti-commutation rules yield:
\begary{rcl}
{\bf b}\,{\bf e}_x\,{\bf b}&=&{\bf e}_x\,{\bf e}_y\,{\bf e}_x\,{\bf e}_x\,{\bf e}_y={\bf e}_x\\
{\bf e}_x\,{\bf b}-{\bf b}\,{\bf e}_x&=&{\bf e}_x\,{\bf e}_x\,{\bf e}_y-{\bf e}_x\,{\bf e}_y\,{\bf e}_x=2\,{\bf e}_y\\
\endary
such that with $c^2-s^2=C$ and $2\,c\,s=S$:
\begary{rcl}
{\bf\tilde x}_x &=&x\,\left({\bf e}_x\,(c^2-s^2)+({\bf e}_y\,2\,c\,s)\right)\\
&=&x\,\left({\bf e}_x\,C+{\bf e}_y\,S\right)\\
\endary
For the $y$-component one obtains equivalently
\begary{rcl}
{\bf\tilde x}_y &=&y\,\left({\bf e}_y\,C-{\bf e}_x\,S\right)\\
\endary
while the $z$-component is unchanged since ${\bf e}_z$ commutes with ${\bf b}_z={\bf e}_x\,{\bf e}_y$. 
In summary we obtain a rotation around the $z$-axis:
\begary{rcl}
{\bf\tilde x}&=&(x\,\cos{(\phi)}-y\,\sin{(\phi)})\,{\bf e}_x\\
&+&(y\,\cos{(\phi)}+x\,\sin{(\phi)})\,{\bf e}_y\,.
\endary
Hence, if such anti-commuting ``unit''-matrices exist, then they can be used to
represent spatial rotations. 

\subsection{Clifford Algebras}

In the previous sections we did not specify the exact form of the matrices ${\bf
  e}_i$ - we only assumed that they exist, mutually anti-commute and square to
the (positive of negative) unit matrix. This means that the exact form of the
matrices is not essential for the purpose of representing rotations. 
This is sometimes interpreted in such a way, that the elements ${\bf e}_i$ do 
not have to be represented by matrices at all. 
Instead it is often suggested to regard ${\bf e}_i$ as abstract elements of a 
so-called Clifford algebra (CA). This view is {\it mathematically} possible
and legitimate, but ignores the intrinsic connection to the concept of {\it
  physical} phase space and the Hamiltonian formalism. Therefore essential 
physical insight, namely the distinction between Hamiltonian and skew-Hamiltonian
elements, is lost.

A Clifford algebra that is generated by three elements ${\bf e}_x$,
${\bf e}_y$ and ${\bf e}_z$ with positive norm (${\bf e}_i^2={\bf 1}$), 
is named $Cl(3,0)$. More generally speaking a Clifford algebra
$Cl(p,q)$ has $N=p+q$ pairwise anti-commuting generators, $p$ of which square to $+{\bf 1}$ and 
$q$ square to $-{\bf 1}$. From combinatorics one finds that $Cl(p,q)$ has $\left(N\atop k\right)$
$k$-vectors and in summary it has
\begeq
\sum\limits_{k=0}^{N-1}\left(N\atop k\right)=2^N
\endeq
linear independent elements, where the $0$-vector is the scalar (unit element)
${\bf 1}$, the {\it vector elements} are the generators of the Clifford algebra
and $k$-vectors are products of $k$ vectors.
The $N$-vector, i.e. the product of all generators, $\prod\limits_{k=0}^{N-1}\,{\bf e}_k$ 
is the so-called {\it pseudo-scalar}. 

$Cl(3,0)$ has $8$ linear independent elements, namely $3$ generators, $3$ bi-vectors 
(Eq.~\ref{eq_rotators}), the scalar ${\bf 1}$ and the pseudoscalar 
${\bf e}_1\,{\bf e}_2\,{\bf e}_3$ (or ${\bf e}_x\,{\bf e}_y\,{\bf e}_z$, respectively).
But since $8$ has no integer root, there is no complete one-to-one relation to a specific 
real square matrix size. A complete one-to-one relation requires that
\begeq
2^N=n^2
\label{eq_uniqueness}
\endeq 
where the matrix size would be $n\times n$. Obviously the condition of completeness 
Eq.~\ref{eq_uniqueness} requires that $N$ is an even number $N=2\,M$. If this is fulfilled,
then 
\begeq
2^{2\,M}=4^M=n^2
\endeq
Then $n^2$ must be a multiple of $4$ so that $n$ must also be even and hence the matrix
dimension is essentially $2\,n\times 2\,n$ and Eq.~\ref{eq_uniqueness} must be
written as
\begeq
2^N=(2\,n)^2
\label{eq_uniqueness2}
\endeq 

However we did not yet consider a time coordinate.
In order to represent a coordinate in Minkowski space-time, a vector has $4$ linear 
independent elements and therefore we introduce another unit element, which might be 
called ${\bf e}_0$ or ${\bf e}_t$. 
Then one has $N=4$ and hence $2^N=16$ linear independent elements, a size that matches to
$4\times 4$-matrices~\footnote{As a result known from representation theory, real squared 
matrices of size $2^m\times 2^m$ can always represent a Clifford algebra, but not all values
of $p$ and $q$ with $p+q=N$ are possible; namely either $p-q=8\,l$ or $p-q=2+8\,l$ 
with arbitary integer $l$ must hold, often written as
\begeq
p-q=0,2\,\,\mathrm{mod}\,\,8\,.
\endeq
This is often called Bott periodicity~\cite{Bott,Okubo}.}.
Real $4\times 4$-matrices allow to represent the Clifford algebras $Cl(2,2)$ and $Cl(3,1)$. 
For our purpose only $Cl(3,1)$ is appropriate, such that ${\bf e}_t^2=-{\bf 1}$. If we refer 
to $4\times 4$-matrices, we use the notation
\begary{rcl}
{\bf e}_t&=&\y_0\\
{\bf e}_x&=&\y_1\\
{\bf e}_y&=&\y_2\\
{\bf e}_z&=&\y_3\\
\endary
A possible choice for the $4$ real $\y$-matrices is given by\footnote{For
  better readability the zeros are replaced by dots.}:
{\small
\begeq
\begin{array}{rclrcl}
\y_0&=&\bmtx{cccc}
   . &   1  &  . &   .\\
  -1 &   .  &  . &   .\\
   . &   .  &  . &   1\\
   . &   .  & -1 &   .\\
\emtx\,,&
 \y_1&=&\bmtx{cccc}
   . &  -1  &  . &   .\\
  -1 &   .  &  . &   .\\
   . &   .  &  . &   1\\
   . &   .  &  1 &   .\\
\emtx\\
 \y_2&=&\bmtx{cccc}
   . &   .  &  . &   1\\
   . &   .  &  1 &   .\\
   . &   1  &  . &   .\\
   1 &   .  &  . &   .\\
\emtx\,,&
 \y_3&=&\bmtx{cccc}
  -1 &   .  &  . &   .\\
   . &   1  &  . &   .\\
   . &   .  & -1 &   .\\
   . &   .  &  . &   1\\
\emtx\\
\end{array}
\label{eq_gammas}
\endeq}
From these $4$ ``generators'' of the Clifford algebra $Cl(3,1)$, which mutually anti-commute, 
the $6$ bi-vectors, the generators of rotations and boosts, are obtained by 
matrix multiplication:
{\small
\begeq
\begin{array}{rclp{4mm}rcl}
\y_4&=&\y_0\,\y_1;&&\y_7&=&\y_2\,\y_3\\
\y_5&=&\y_0\,\y_2;&&\y_8&=&\y_3\,\y_1\\
\y_6&=&\y_0\,\y_3;&&\y_9&=&\y_1\,\y_2\\
\end{array}
\endeq}
Hence the matrices $\y_7$, $\y_8$ and $\y_9$ represent the bi-vector ${\bf b}$ of
Eq.~\ref{eq_rotators}. Since the new generator $\y_0$ anti-commutes with $\y_1$, $\y_2$ 
and $\y_3$, it commutes with $\y_7$, $\y_8$ and $\y_9$ and is hence unchanged by the 
rotations generated by (the matrix exponential of) these bi-vectors. It is therefore
no spatial coordinate.
Furthermore we have $3$ more bi-vectors $\y_4$, $\y_5$ and $\y_6$, which
square to $+{\bf 1}$:
\begeq
\y_4^2=(\y_0\,\y_1)^2=-\y_0^2\,\y_1^2={\bf 1}\,.
\endeq
From Eq.~\ref{eq_boostgen} we know that $\y_4$, $\y_5$ and $\y_6$ generate boosts,
not rotations. As ${\bf b}_z=\y_9=\y_1\,\y_2={\bf e}_x\,{\bf e}_y$ 
generates rotations in the $x-y$-plane, the bi-vector $\y_4=\y_0\,\y_3$ generates a 
boost in the ``plane'' of $\y_0$ and $\y_3$.

\subsection{Boosts as Similarity Transformations}
\label{sec_boost2}

We now examine the result of the transformation of a ``vector'' ${\bf x}=t\,\y_0+x\,\y_1+ y\,\y_2+z\,\y_3$
in more detail:
\begeq
{\bf\tilde x}={\bf B}\,{\bf x}\,{\bf B}^{-1}\,,
\endeq
where ${\bf B}=\exp{(-\y_0\,\y_3\,\tau/2)}$. The product $\y_0\,\y_3$ commutes
with both $\y_1$ and $\y_2$, so that $\tilde x=x$ and $\tilde y=y$. For the
other two components we evaluate component-wise~\footnote{
Given an arbitrary matrix ${\bf F}=\sum_k\,f_k\,\y_k$ that is an unknown vector. 
Since the trace of all Dirac matrices vanishes except for the unit matrix,
one obtains the coefficient $f_k$ of $\y_k$ by the formula
\begeq
f_k=\frac{1}{4}\,\mathrm{Tr}(\y_k^T\,{\bf F})
\endeq
} with $c\equiv\cosh{(\tau/2)}$ and $s\equiv\sinh{(\tau/2)}$:
\begary{rcl}
\tilde t\,\y_0+\tilde z\,\y_3&=&({\bf 1}\,c-\y_0\,\y_3\,s)\,(t\,\y_0+z\,\y_3)\,({\bf 1}\,c+\y_0\,\y_3\,s)\\
&=&t\,\left(\y_0\,(c^2+s^2)-2\,c\,s\,\y_3\right)\\
&+&z\,\left(\y_3\,(c^2+s^2)-2\,c\,s\,\y_0\right)\\
&=&t\,(\y_0\,C-S\,\y_3)+z\,(\y_3\,C-S\,\y_0)\\
&=&\y_0\,(t\,C-z\,S)+\y_3\,(z\,C-t\,S)\,,
\endary
where with $C=\cosh{(\tau)}$ and $S=\sinh{(\tau)}$, 
we used the following theorems 
\begary{rcl}
\cosh^2{(\tau/2)}+\sinh^2{(\tau/2)}&=&\cosh{(\tau)}\\
2\,\cosh{(\tau/2)}\,\sinh{(\tau/2)}&=&\sinh{(\tau)}\,.
\endary
If we use the conventional notation $\y=\cosh{(\tau)}$ and 
$\beta=\tanh{(\tau)}$ (i.e. $\beta\,\y=\sinh{(\tau)}$), then we obtain the Lorentz boost
along the $z$-axis
\begary{rcl}
\tilde t&=&\y\,t-\beta\y\,z\\
\tilde z&=&\y\,z-\beta\y\,t\\
\endary
where $\tau=\mathrm{artanh}{(\beta)}$ is the so-called ``rapidity''.

Thus we have demonstrated that a $4$-vector in Minkowski space-time has a
natural representation by matrices and that both, rotations and boosts of
$4$-vectors, can be written as similarity transformations. Next we prove
that rotations and boosts of electromagnetic fields follow the exact same
approach, i.e. can be represented by exactly the same 
similarity transformations, if the fields are ``encoded'' as bi-vectors:
\begary{rcl}
\vec E&\to&\y_0\,\vec E\cdot\vec\y\equiv E_x\,\y_4+E_y\,\y_5+E_z\,\y_6\\
\vec B&\to&\y_{14}\,\y_0\,\vec B\cdot\vec\y\equiv B_x\,\y_7+B_y\,\y_8+B_z\,\y_9\\
\endary
with the pseudo-scalar $\y_{14}=\y_0\,\y_1\,\y_2\,\y_3$.

\subsection{Rotations of Electromagnetic fields}
\label{sec_emrot}

Again we use a rotation around the $z$-axis (see Eq.~\ref{eq_rot}), i.e.
the generator is $\y_9=\y_1\,\y_2$ and it commutes with $\y_9$, which is trivial
and with $\y_6=\y_0\,\y_3$, which is also quickly verified. But $ \y_9$ anti-commutes
with $\y_4=\y_0\,\y_1$ and $\y_5=\y_0\,\y_2$, so that:
\begary{rcl}
\tilde E_z&=&E_z\\
\tilde B_z&=&B_z\\
\endary
The electric field components in the $x-y$-plane are 
(with $c=\cos{(\phi/2)}$ and $s=\sin{(\phi/2)}$, $C=\cos{(\phi)}$ and $S=\sin{(\phi)}$):
\begary{rcl}
\tilde E_x\,\y_4+\tilde E_y\,\y_5&=&(c-s\,\y_1\,\y_2)\,(E_x\,\y_4+E_y\,\y_5)\,(c+s\,\y_1\,\y_2)\\
&=&E_x\,(\y_4\,(c^2-s^2)+2\,s\,c\,\y_5)\\
&+&E_y\,(\y_5\,(c^2-s^2)-2\,s\,c\,\y_4)\\
&=&E_x\,(\y_4\,C+S\,\y_5)+E_y\,(\y_5\,C-S\,\y_4)\\
&=&\y_4\,(E_x\,C-E_y\,S)+\y_5\,(E_y\,C+E_x\,S)\\
\tilde E_x&=&E_x\,\cos{(\phi)}-E_y\,\sin{(\phi)}\\
\tilde E_y&=&E_x\,\sin{(\phi)}+E_y\,\cos{(\phi)}\\
\endary
The terms of the magnetic field transform in exactly the same way:
\begary{rcl}
\tilde B_x\,\y_7+\tilde B_y\,\y_8&=&(c-s\,\y_9)\,(B_x\,\y_7+B_y\,\y_8)\,(c+s\,\y_9)\\
&=&B_x\,(\y_7\,(c^2-s^2)+2\,s\,c\,\y_8)\\
&+&B_y\,(\y_8\,(c^2-s^2)-2\,s\,c\,\y_7)\\
&=&B_x\,(\y_7\,C+S\,\y_8)+B_y\,(\y_8\,C-S\,\y_7)\\
&=&\y_7\,(B_x\,C-B_y\,S)+\y_8\,(B_y\,C+B_x\,S)\\
\tilde B_x&=&B_x\,\cos{(\phi)}-B_y\,\sin{(\phi)}\\
\tilde B_y&=&B_y\,\sin{(\phi)}+B_x\,\cos{(\phi)}\\
\endary

\subsection{Boosts of Electromagnetic fields}
\label{sec_emboost}

A boost along $z$ is generated by $\y_6=\y_0\,\y_3$, which commutes with itself and
with $\y_9$, such that the electromagnetic field components in the direction of the boost
are unchanged. The electric field components in the plane perpendicular to the boost are
(with $c=\cosh{(\tau/2)}$ and $s=\sinh{(\tau/2)}$, $C=\cosh{(\tau)}$ and $S=\sinh{(\tau)}$):
\begary{rcl}
\tilde E_x\,\y_4+\tilde E_y\,\y_5&=&(c-s\,\y_6)\,(E_x\,\y_4+E_y\,\y_5)\,(c+s\,\y_6)\\
&=&E_x\,(\y_4\,(c^2+s^2)-s\,c\,\y_6\,\y_4+s\,c\,\y_4\,\y_6)\\
&+&E_y\,(\y_5\,(c^2+s^2)-s\,c\,\y_6\,\y_5+s\,c\,\y_5\,\y_6)\\
\endary
With $\y_4\,\y_6=\y_0\,\y_1\,\y_0\,\y_3=\y_1\,\y_3=-\y_8$ and 
$\y_5\,\y_6=\y_0\,\y_2\,\y_0\,\y_3=\y_2\,\y_3=\y_7$ we obtain: 
\begary{rcl}
(c-s\,\y_6)\,E_x\,\y_4\,(c+s\,\y_6)&=&E_x\,(\y_4\,C-S\,\y_8)\\
(c-s\,\y_6)\,E_y\,\y_5\,(c+s\,\y_6)&=&E_y\,(\y_5\,C+S\,\y_7)\\
\endary
With $\y_6\,\y_7=\y_0\,\y_3\,\y_2\,\y_3=-\y_0\,\y_2=-\y_5$ and 
$\y_6\,\y_8=\y_0\,\y_3\,\y_3\,\y_1=\y_0\,\y_1=\y_4$ we obtain: 
\begary{rcl}
(c-s\,\y_6)\,B_x\,\y_7\,(c+s\,\y_6)&=&B_x\,(\y_7\,C+S\,\y_5)\\
(c-s\,\y_6)\,B_y\,\y_8\,(c+s\,\y_6)&=&B_y\,(\y_8\,C-S\,\y_4)\\
\endary
such that (again with $C=\y$ and $S=\beta\y$):
\begary{rcl}
\tilde E_x&=&\y\,E_x-\beta\y \,B_y\\
\tilde E_y&=&\y\,E_y+\beta\y \,B_x\\
\tilde B_x&=&\y\,B_x+\beta\y \,E_y\\
\tilde B_y&=&\y\,B_y-\beta\y \,E_x\\
\endary
These equations are in exact agreement with the Lorentz transformation of the
electromagnetic fields.

\subsection{The Lorentz Force}

Hence we obtain a perfectly simple and systematic approach not only of rotations
but also of boosts, if we associate the $4$-vector components with $\y_0$ (time-like,
energy ${\cal E}$) and $\y_1$, $\y_2$ and $\y_3$ for the space-like components (momentum, $\vec P$)
and furthermore associate electromagnetic fields with the bi-vectors~\footnote{
This mapping has been called electro-mechanical equivalence (EMEQ)~\cite{rdm_paper,geo_paper}.}:
\begary{rcl}
{\cal E}&\to&{\cal E}\,\y_0\\
\vec P&\to&P_x\,\y_1+P_y\,\y_2+P_z\,\y_3\\
\vec E&\to&E_x\,\y_4+E_y\,\y_5+E_z\,\y_6\\
      &=&\y_0\,(E_x\,\y_1+E_y\,\y_2+E_z\,\y_3)\\
\vec B&\to&B_x\,\y_7+B_y\,\y_8+B_z\,\y_9\\
      &=&B_x\,\y_2\,\y_3+B_y\,\y_3\,\y_1+B_z\,\y_1\,\y_2\\
\label{eq_EMEQ}
\endary
This mapping has {\it physical significance} firstly, because magnetic fields
{\it actively act} as generators of rotational motion (in momentum space)
and electric fields {\it actively act} as generators of boosts (of charged particles), 
and secondly, with the use of the appropriate scaling factor ${q\over 2\,m}$,
the Lorentz force can be written as~\cite{rdm_paper,geo_paper}:
\begary{rcl}
{\bf P}&=&{\cal E}\,\y_0+P_x\,\y_1+P_y\,\y_2+P_z\,\y_3\\
{\bf F}&=&E_x\,\y_4+E_y\,\y_5+E_z\,\y_6+B_x\,\y_7+B_y\,\y_8+B_z\,\y_9\\
{\bf \dot P}&=&{q\over 2\,m}\,({\bf F}\,{\bf P}-{\bf P}\,{\bf F})\\
\label{eq_LorentzForce0}
\endary
where the overdot indicates the derivative with respect to proper time.
$q$ and $m$ are charge and mass of the particle and are required to obtain
electric and magnetic field in the units of frequency.
The evaluation of the components gives, translated back into conventional 
vector form:
\begary{rcl}
\dot{\cal E}&=&{q\over m}\,\vec P\cdot\vec E\\
\dot{\vec P}&=&{q\over m}\,({\cal E}\,\vec E+\vec P\times\vec B)\\
\endary
with $d\tau=dt/\y$ this becomes (with $c=1$):
\begary{rcl}
{d{\cal E}\over dt}&=&{q\over m\,\y}\,\vec P\cdot\vec E=q\,\vec v\cdot\vec E\\
{d\vec P\over dt}&=&q\,\vec E+q\,\vec v\times\vec B\\
\label{eq_LorentzForce1}
\endary
To summarize: if we make use of ten {\it Hamiltonian} elements (out of $16$) of the Clifford algebra
$Cl(3,1)$, we find a systematic description of minimal complexity for a massive particle
in an (``external'') electromagnetic field - simply by the use of $4\times 4$-matrices
instead of the conventional vector-notation. The idea to use real unit {\it
  matrices} instead of unit {\it vectors} thus lead us directly to the
structure of Minkowski space-time, i.e. to the ``real physical space''.

How is this possible and what about the remaining six elements of the complete
Clifford algebra?

\subsection{The Remaining Matrices}
\label{sec_remain}

The remaining $6$ matrices are not directly used, but are given to complete the list 
of $16$ real $\y$-matrices:
{\small
\begeq
\begin{array}{rclp{4mm}rcl}
\y_{14}&=&\y_0\,\y_1\,\y_2\,\y_3;&&\y_{15}&=&{\bf 1}\\
\y_{10}&=&\y_{14}\,\y_0&=&\y_1\,\y_2\,\y_3&&\\
\y_{11}&=&\y_{14}\,\y_1&=&\y_0\,\y_2\,\y_3&&\\
\y_{12}&=&\y_{14}\,\y_2&=&\y_0\,\y_3\,\y_1&&\\
\y_{13}&=&\y_{14}\,\y_3&=&\y_0\,\y_1\,\y_2&&\\
\end{array}
\endeq}
where $\y_{14}$ is the pseudoscalar, $\y_{15}$ the unit matrix and
the matrices $\y_{10}$ up to $\y_{13}$ are so-called axial vectors.

As we have shown above, all LTs (rotations and boosts) can be written
in the general form of a similarity transformation (Eq.~\ref{eq_LT}), if
Eq.~\ref{eq_EMEQ} is used to compose the matrix ${\bf F}$: $4$-vectors
$(u_0,{\bf u})$ enter the matrix ${\bf F}$ as coefficients of the
$\y_\mu$-matrices and ``tensor'' components as coefficients of the
corresponding bi-vectors. Raising and lowering of indices is then obsolete.

As we have shown, the essence of relativistic kinematics, namely the Lorentz 
transformations of both, 4-vectors and electromagnetic fields, matches the 
Clifford algebraic decomposition of real $4\times 4$-matrices. 
But why is this so, why do we need a matrix exponential, how do we arrive
at Eq.~\ref{eq_LorentzForce0} and why do we use only $10$ out of $16$ matrices?
And, since we use the Dirac algebra: is all this related to the Dirac equation
and if so, why don't we need to use complex numbers? As we will show in the next 
section, all of these questions can be answered on the basis of Hamiltonian theory.

\section{Phase Space}
\label{sec_phasespace}

Goldstein's ``Classical Mechanics'' contains the following statement:
``The advantages of the Hamiltonian 
formulation lie not in its use as a calculational tool, but rather in the deeper 
insight it affords into the formal structure of mechanics. The equal status accorded 
to coordinates and momenta as independent variables encourages a greater  freedom 
in selecting the physical quantities to be designated as "coordinates" and 
"momenta." As a result we are led to newer, more abstract ways of presenting 
the physical content of mechanics. While often of considerable help in practical 
applications to mechanical problems, these more abstract formulations are primarily
of interest to us today because of their essential role in constructing the more 
modern theories of matter.''~\cite{Goldstein}.

We suggest in this article to make use of the mentioned freedom, and to replace
the conventional relation of phase space points and measurable quantities by something
more abstract: While the naive realist take of classical physics narrows the possible 
meaning of a phase space point to the spatial position and mechanical momentum of a 
mass point, quantum mechanics can most naturally be understood by the use of an 
indirect relation. It has been suggested that this indirect relation is 
a statistical one, namely that the measurable quantities listed in Eq.~\ref{eq_EMEQ}, 
are (second) {\it moments} in phase space~\cite{GLT,qed_paper,osc_paper}.
According to this view, spinors are (ensembles of) points in an abstract phase space
underlying both special relativity and quantum mechanics. A ``particle'' is then
represented by a classical Hamiltonian ensemble.

\subsection{The Hamiltonian}
\label{sec_hamilton}

The structure of the Dirac algebra has for instance been described by Albert 
Messiah~\cite{Messiah}, the geometric content of which has been described
by Lounesto and Hestenes~\cite{Lounesto,STA}. Our account differs from the 
conventional form by the 
use of the metric $g=\mathrm{Diag}(-1,1,1,1)$, i.e. $\y_0^2=-{\bf 1}$ and
$\y_k^2={\bf 1}$ for $k\in[1,2,3]$. The motivation for the use of a different
metric and of the {\it real} Dirac matrices instead of the conventional
complex form is, besides the reduction of complexity, that the Clifford algebra 
$Cl(3,1)$ can be derived from a general quadratic Hamiltonian of two classical DOF.
Hence $Cl(3,1)$ provides the toolbox to describe arbitrary linear couplings of two DOF 
and therefore has a fundamental algebraic {\it and} physical significance.
This is not limited to the Dirac equation, not even to quantum mechanics: 
It is a general and fundamental algebraic tool in Hamiltonian phase 
space~\cite{rdm_paper,geo_paper,jacobi}. 

The algebra $Cl(3,1)$ includes all Hamiltonian generators $sp(4)$ of linear 
canonical transformations of two degrees of freedom~\footnote{
This means, that we follow a hint given by Res Jost and mentioned at the end 
of Dirac's celebrated paper on the $3+2$ de Sitter Group~\cite{Dirac}, namely 
the connection of the Dirac algebra with the Lie algebra $sp(4)$ of the real 
symplectic group $Sp(4)$.}.
It has been emphasized by several authors that the complex wave-function can be 
transformed into a ``classical'' Hamiltonian phase space point~\cite{Strocchi,GV,Ralston,Khrennikov,Briggs}. 
Accordingly one can derive major aspects of quantum mechanics from 
classical Hamiltonian concepts.

One may recall Kepler's reasoning: Simplifying the math as a path towards
physical insight. Kepler did not know the physical reason behind his laws
(i.e. gravitation), but the remarkable conceptual simplification 
of the description of planetary orbits by his laws provided the
ground for the formulation of Newton's law of gravitation.

Indeed it has been suggested that ``the quantum paradoxes of Bell, Kochen and Specker,
Greenberger et al. and Hardy can be formally considered from a single view-point: 
they are all examples of the failure to find a solution to a certain moments' problem''
~\cite{Klyshko}.

In two preceeding essays we argued that, on some fundamental level, dynamical
variables (DV) can not be directly observable. Only the second and higher (even)
moments of the DV are direct observables~\cite{qed_paper,fit_paper}.
How this has to be understood will be explained in the following~\footnote{
Note that our account of the Lorentz transformations is fully equivalent to 
that of the Dirac spinor as used in conventional QED. Schm\"user has given
a relatively clear and simple account, albeit using the complex version
of the Dirac matrices with metric $(1,-1,-1,-1)$~\cite{Schmueser}.}.

Let $\psi$ be a phase space point $\psi=(q_1,p_1,q_2,p_2)^T$ of a system with two degrees
of freedom, where $q_i$ and $p_i$ represent {\it unspecified} dynamical variables.
The most general form for a non-singular Hamiltonian function of two degrees of
freedom can be expressed by a Taylor series in four variables. If we cut the
Taylor series after the second order terms, this approach is equivalent to
a theory of small oscillations.

The general second-order Hamiltonian function of a two ``classical'' DOF
is given by~\cite{qed_paper,osc_paper,fit_paper}:
\begeq
{\cal H}(\psi)=\frac{1}{2}\,\psi^T\,{\bf A}\,\psi\,,
\label{eq_Hamilton}
\endeq
We assume that ${\bf A}$ can be an arbitrary {\it symmetric} real
$4\times 4$ matrix. The Hamiltonian equations of motion then yield:
\begeq
\dot\psi=\y_0\,{\bf A}\,\psi={\bf F}\,\psi
\label{eq_eqom}
\endeq
$\y_0$ is a $4\times 4$ symplectic unit matrix (SUM), which means that it is 
skew-symmetric and orthogonal such that $\y_0^2=-{\bf 1}$ and represents with
this properties the structure of the Hamiltonian equations of motion. 
The chosen form (Eq.~\ref{eq_gammas}) complies with the order of the abstract 
phase space coordinates $q_i$ and $p_i$ in $\psi$ and the notational
convention of the Hamiltonian equations of motion
\begary{rcl}
\dot q_i&=&{\d{\cal H}\over\d p_i}\\
\dot p_i&=&-{\d{\cal H}\over\d q_i}\,,
\label{eq_Heqom}
\endary
which means that Eq.~\ref{eq_eqom} is the result of inserting Eq.~\ref{eq_Hamilton}
into Eq.~\ref{eq_Heqom}.

\subsection{Hamiltonian Algebra}

The theory of symplectic motion, as it is usually presented, suffers
from over-{\it geometrization}. One can not resist the impression that
theorist are fixated with geometry, almost completely leaving aside the 
fundamental {\it temporal}, {\it algebraic}, and {\it statistical} aspects of the 
notion of Hamiltonian {\it phase space}. It is also remarkable, that, while 
it is widely supported that the mysterious features of quantum mechanics should 
be taught in secondary school, the notion of a phase space, which is central 
to almost every branch of physics and an {\it inevitable notion in QM}, is 
sometimes not taught at all or just briefly mentioned -- as if it was 
somehow dispensable. Similarily, the Dirac equation 
is almost {\it banned} from curricula, often just briefly discussed 
in the second volume of quantum mechanics textbooks and rarely ever 
mentioned in discussions concerning the interpretation of quantum mechanics.
As Hestenes remarked, ``[it] has long puzzled me is why Dirac 
theory is almost universally ignored in studies on the interpretation of quantum
mechanics, despite the fact that the Dirac equation is widely recognized
as the most fundamental equation in quantum mechanics''~\cite{Hestenes}.

We believe that, once properly understood, the connection of the Dirac
equation to the notion of a classical phase space has a unique potential 
to provide deeper insights into the mathematical principles of physics, while 
being itself simple, clear and straightforward.

A matrix ${\bf S}$ is said to be Hamiltonian, if it obeys~\cite{MHO}
\begeq
 {\bf S}^T=\y_0\,{\bf S}\,\y_0
\label{eq_symplex}
\endeq
and a matrix ${\bf C}$ is said to be {\it skew-Hamiltonian}, if it obeys
\begeq
 {\bf C}^T=-\y_0\,{\bf C}\,\y_0
\label{eq_cosymplex}
\endeq
The meaning of this distinction is simply the follwoing: Hamiltonian
matrices are similar to $\y_0\,{\bf A}$, i.e. they are exclusively
  composed of terms that may appear in a Hamiltonian function and are
  therefore possible generators of canonical transformations, while
  the contribution of {\it skew-Hamiltonian} matrices to the Hamiltonian function
  vanishes.

It is easy to prove that $\y_0\,{\bf S}$ is symmetric and $\y_0\,{\bf C}$ is
skew-symmetric. The interesting point to note here is that the Hamiltonian
structure, as represented by $\y_0$, connects matrix symmetries (concerning 
transposition) with commutativity. If in Eq.~\ref{eq_symplex} the matrices 
$\y_0$ and ${\bf S}$ commute, then 
\begeq
 {\bf S}^T=\y_0^2\,{\bf S}=-{\bf S}
\endeq
and hence ${\bf S}$ must be skew-symmetric. Indeed the Hamiltonian formalism
generates the algebraic properties of Eq.~\ref{eq_cosy_algebra} given below.

Hence the matrix ${\bf F}$ is Hamiltonian and since it is
the product of a symmetric and a skew-symmetric matrix, the trace vanishes:
\begeq
\mathrm{Tr}({\bf F})=0\,.
\endeq

Any real symmetric $4\times 4$ matrix ${\bf A}$ (Eq.~\ref{eq_Hamilton}) has 
ten linear independent real parameters~\footnote{See also Ref.~(\cite{Dirac,Kim}).},
and the same holds for ${\bf F}$. 
The solution of Eq.~\ref{eq_eqom}, for constant ${\bf F}$, is given
by the matrix exponential of ${\bf F}$:
\begeq
\psi(\tau)=\exp{({\bf F}\,\tau)}\,\psi(0)={\bf M}(\tau)\,\psi(0)\,.
\endeq
The matrix exponential of a Hamiltonian matrix (see below) is a symplectic
matrix, i.e. a canonical transformation~\cite{MHO}. Since any exponential of a 
Hamiltonian matrix is symplectic, and since all driving terms of the Lorentz
transformations are (in this approach) Hamiltonian matrices, the Lorentz 
transformations are symplectic similarity transformations that can be derived 
from the Hamiltonian function of two classical (coupled) DOF.
The eigenvalues of the Hamiltonian matrix are constants of motion, since
all possible (Lorentz-) transformations are similarity transformations.
In App.~\ref{app_eigen} we show that the eigenvalues in an inertial system
are identical to the mass such that the mass is Lorentz invariant.

A matrix ${\bf S}$ is said to be symplectic, if it obeys
\begeq
 {\bf S}^T\,\y_0\,{\bf S}=\y_0
\label{eq_symplectic}
\endeq
and a matrix ${\bf C}$ is said to be cosymplectic~\footnote{
Elsewhere it would be called symplectic with multiplyer $-1$~\cite{MHO}. 
}, if it obeys
\begeq
 {\bf C}^T\,\y_0\,{\bf C}=-\y_0
\label{eq_cosymplectic}
\endeq
Since the equations of motion Eq.~\ref{eq_eqom} contain, by definition,
only Hamiltonian terms, cosymplectic transformations can not be derived
from a non-zero Hamiltonian function.

If we presume (or argue~\cite{qed_paper}) that, on this level of description,
observables are always (averaged) amplitudes and never phases, then the
observables are (derived from) second (and higher even) moments of a density
distribution of phase space points $\rho(\psi)$.
As in classical statistical mechanics, we can likewise think of a
particle density or of the probability to find a system in a certain state. 
The suggested second order Hamiltonian function, integrated over the density, 
is then proportional to a linear combination of second moments of the 
phase space density.

\subsection{Second Moments in Phase Space}

The second moments in phase space form a matrix $\Sigma$:
\begeq
\Sigma_{ij}=\langle(\psi_i-\langle\psi_i\rangle)(\psi_j-\langle\psi_j\rangle)\rangle
\endeq
where the angles indicate the phase space average.
The first moments either vanish or can be made to vanish by an appropriate
choice of the origin, so that the second moments are
\begeq
\Sigma_{ij}=\langle \psi_i\psi_j\rangle=\langle \psi\psi^T\rangle\,.
\endeq
The time evolution of the second moments is then obtained by inserting 
Eq.~\ref{eq_eqom}:
\begary{rcl}
\dot\Sigma&=&\langle\dot\psi\psi^T\rangle+\langle \psi\dot\psi^T\rangle\\
      &=&{\bf F}\,\langle\psi\psi^T\rangle+\langle \psi\psi^T\rangle\,{\bf F}^T\\
      &=&{\bf F}\,\Sigma+\Sigma\,\y_0\,{\bf F}\,\y_0\\
\endary
so that by multiplication with $\y_0^T$ from the right one obtains~\footnote{
These equations are often called {\it envelope equations}, for instance in accelerator
physics, where the (roots of the) second moments of the beam phase space distribution
are used to provide a measure of the size of a beam envelope.
}:
\begary{rcl}
\dot\Sigma\y_0^T&=&{\bf F}\,\Sigma\y_0^T-\Sigma\,\y_0^T\,{\bf F}\\
{\bf\dot S}&=&{\bf F}\,{\bf S}-{\bf S}\,{\bf F}\\
\label{eq_envelope}
\endary
where 
\begeq
{\bf S}\equiv\Sigma\y_0^T
\label{eq_Smtx}
\endeq
$\y_0=-\y_0^T$ and $\y_0\y_0^T={\bf 1}$.
Note that Eq.~\ref{eq_envelope} and Eq.~\ref{eq_LorentzForce0} have the exact
same form.
It follows from Eq.~\ref{eq_envelope} that a stable situation ${\bf\dot S}=0$
implies commuting matrices. Commuting matrices share a system of eigenvectors.
Hence eigenvectors and eigenvalues are necessary (or at least adequate) to
describe classical oscillatory motion and are no inventions of quantum physics.

The matrix ${\bf S}$ is, like ${\bf F}$, a {\it Hamiltonian} matrix and
can be written as a product of a symmetric matrix and the SUM $\y_0$.
Unfortunately, the notion of the {\it Hamiltonian matrix}, has also been used
differently by physicists, for instance by Feynman~\cite{Feynman}. Therefore it has 
been suggested to use a different naming convention, borrowed from ``symplectic'' and ``complex'', 
according to which a Hamiltonian matrix ${\bf S}$ that holds Eq.~\ref{eq_symplex} is 
called {\it symplex} (plural {\it symplices}) and a skew-Hamiltonian matrix that 
holds Eq.~\ref{eq_cosymplex} is called {\it cosymplex}~\cite{qed_paper,geo_paper,osc_paper}.
The equations of motion (Eq.~\ref{eq_eqom}) derived from the Hamiltonian, 
are driven by a ``symplex'' ${\bf F}$: Only symplices represent non-zero expectation values,
since all basic expectation values are elements of the auto-correlation matrix $\Sigma$.
Cosymplices have vanishing expectation values and may not appear as driving terms in
linear Hamiltonian theory. As we have shown in Ref.~\cite{qed_paper}, the distinction
between Hamiltonian and skew-Hamiltonian terms (i.e. symplices and cosymplices) allows
to derive the Maxwell equations and this approach explains why magnetic monopoles
don't exist. 

Furthermore Eq.~\ref{eq_envelope} establishes a Lax pair~\cite{Lax}, namely ${\bf S}$
and ${\bf F}$ so that the trace of any power of ${\bf S}$ is a constant of motion:
\begeq
\mathrm{Tr}({\bf S}^k)=\mathrm{const}
\endeq
It will be shown in the next section that odd exponents ${\bf S}^{2m+1}$ are 
again Hamiltonian. This implies that odd exponents have vanishing trace. 
Only for even $k$ the expression yields non-vanishing ``constants of motion'':
\begeq
\mathrm{Tr}({\bf S}^{2k})=\mathrm{const}
\endeq

\subsection{Hamiltonian Clifford Algebras}

The symplectic unit matrix $\y_0$ itself is a symplex (i.e. Hamiltonian):
\begeq
\y_0^T=\y_0^3=-\y_0
\endeq
If a symplex $\y_k\ne\y_0$ anticommutes with $\y_0$, then its matrix
representation is symmetric:
\begary{rcl}
\y_k^T&=&\y_0\,\y_k\,\y_0\\
      &=&-\y_0\,\y_0\,\y_k\\
      &=&\y_k\,,
\endary
since $\y_0^2=-{\bf 1}$. It follows that all generators of $Cl(3,1)$ are
symplices, i.e. driving terms of the Hamiltonian, while in $Cl(2,2)$
at least one generator can not appear in the Hamiltonian: If a Clifford
algebra has $q$ skew-symmetric generators, one of them being the SUM $\y_0$, 
then $q-1$ generators are cosymplices (skew-Hamiltonian). This means that with
respect to the possibility to represent space-time coordinates, the condition
that the generators of the Clifford algebra are symplices (that they {\it can}
contribute to the Hamiltonian), selects space-times with a single generator
associated with time (or energy, respectively). 

For any Hamiltonian system of size $2\,n\times 2\,n$ we find that, if
${\bf S}$ denotes a symplex and ${\bf C}$ a cosymplex, then the 
following rules for (anti-) commutators are obtained: 
\begary{ccc}
\left.\begin{array}{c}
{\bf S}_1\,{\bf S}_2-{\bf S}_2\,{\bf S}_1\\
{\bf C}_1\,{\bf C}_2-{\bf C}_2\,{\bf C}_1\\
{\bf C}\,{\bf S}+{\bf S}\,{\bf C}\\
{\bf S}^{2\,n+1}\\
\end{array}\right\} & \Rightarrow & \mathrm{symplex}\\&&\\
\left.\begin{array}{c}
{\bf S}_1\,{\bf S}_2+{\bf S}_2\,{\bf S}_1\\
{\bf C}_1\,{\bf C}_2+{\bf C}_2\,{\bf C}_1\\
{\bf C}\,{\bf S}-{\bf S}\,{\bf C}\\
{\bf S}^{2\,n}\\
{\bf C}^n\\
\end{array}\right\} & \Rightarrow & \mathrm{cosymplex}\\
\label{eq_cosy_algebra}
\endary
If, as in case of $n=1$ and $n=2$, the algebra is not only 
Hamiltonian, but also a Clifford algebra, then it is appropriate
to identify the ${\bf S}_i$ and ${\bf C}_j$ with the elements
of the Clifford algebra such that any combination of ${\bf S}_i$ 
and ${\bf C}_j$ either commute or anti-commute.
Then it is also easily shown that all basic elements of the 
algebra (all $\y_k$) are either  a symplex or a cosymplex, 
either symplectic or cosymplectic and either symmetric or 
skew-symmetric. In this case we speak of a {\it Hamiltonian
Clifford Algebra} (HCA).

\begin{table}
\begin{tabular}{|c||c|c|c|c||}\hline
Type         & Elements & Order $k$ & c/s & Elements \\\hline
Scalar       & 1        & 0     & c   & ${\bf 1}$\\
Vector       & 1+3=4    & 1     & s   & $\y_0$,($\y_1$,$\y_2$,$\y_3$)\\
Bi-Vector    & 3+3=6    & 2     & s   & ($\y_4$,$\y_5$,$\y_6$),($\y_7$,$\y_8$,$\y_9$)\\
3-Vector     & 1+3=4    & 3     & c   & $\y_{10}$,($\y_{11}$,$\y_{12}$,$\y_{13}$)\\
Pseudoscalar & 1        & 4     & c   & $\y_{14}$\\\hline
\end{tabular}
\caption[]{
  \label{tab_order}
  The elements of the Hamiltonian Clifford algebra $Cl(3,1)$ (real Dirac algebra).
  The column labeled ``c/s'' indicates (s)ymplices and (c)osymplices.
}
\end{table}
Now it is a arguably a physical requirement that all generators of the HCA 
must be Hamiltonian: Since any $k$-vector of the HCA is a 
product of $k$ symplices ${\bf S}_1\dots{\bf S}_k$, one finds that 
(where ${\bf S}_i$ is some generator of the HCA):
\begary{rcl}
({\bf S}_1\,{\bf S}_2\,\dots\,{\bf S}_k)^T&=&{\bf S}_k^T\,{\bf S}_{k-1}^T\,\dots\,{\bf S}_1^T\\
&=&\y_0\,{\bf S}_k\,\y_0^2\,{\bf S}_{k-1}\,\y_0^2\,\dots\,\y_0^2\,{\bf S}_1\,\y_0\\
&=&(-1)^s\,\y_0\,{\bf S}_k\,{\bf S}_{k-1}\,\dots\,{\bf S}_1\,\y_0\\
&=&(-1)^t\,\y_0\,{\bf S}_1\,{\bf S}_2\,\dots\,{\bf S}_k\,\y_0\\
\endary
where $s=k-1$ from the number of factors $\y_0^2=-{\bf 1}$ (third to fourth
row), while $t=s+a$ where $a$ is the number of commutations required to
reverse the order of $k$ anti-commuting elements, given by combinatorics as 
$a=k\,(k-1)/2$. Hence we find that such
$k$-vectors are symplices, if $t=k-1+k\,(k-1)/2=k/2-1+k^2/2$ is even. This is
the case for 
\begeq
k=1,2,5,6,9,10,\dots
\label{eq_order}
\endeq
It is surprizing and remarkable that this kind of periodicity appears, since it
shows that possible types of interactions (transformations) have narrow algebraic
constraints. Since the highest vector order $k$ of $Cl(p,q)$ is $k\le N=p+q$, then
in the algebra $Cl(3,1)$ the value $k$ is constrained to $0\le k\le 4$, so that all 
symplices are either vectors ($k=1$) or bi-vectors ($k=2$), i.e. exactly the elements 
of Eq.~\ref{eq_EMEQ}, so that 
\begeq
{\bf F}={\cal E}\,\y_0+{\vec p}\cdot\vec\y+\y_0\,\vec E\,\cdot\vec\y+\y_{14}\,\y_0\,\vec B\cdot\,\vec\y\,.
\endeq
where $\vec p$ is, as we take from the equal form of Eq.~\ref{eq_LorentzForce0}
and Eq.~\ref{eq_envelope}, the {\it mechanical} momentum.

\subsection{Observables are Generators are Observables}

It is a fundamental finding of classical physics that the driving terms of change
(the generators or Lorentz transformations, for instance) are themselves {\it observable}
and vice versa: Energy is the generator of time-translations, the momentum is the generator 
of spatial translations, the angular momentum is the generator of rotations and so on. 
This kind of closure has widely been ignored in textbook treatments of the Lorentz 
transformations: the algebraic terms that generate boosts are related to electric fields 
and those that generate rotations are related to gyroscopic quantities like spin, 
angular momentum or magnetic field. 

Lorentz transformations are most often treated as coordinate transformations in space-time
{\it without} any detailed analysis of how these transformations are generated. However
neither a coordinate system nor a coordinate transformation are {\it per se} physical, 
{\it unless} one finds the generators and observables of these transformations in the context
of a dynamical theory. The conventional treatment starts from a quasi-Newtonian perspective,
i.e. from the apriori assumption of some self-sufficient space-time that imposes constraints
on possible dynamics. Here we suggest to reverse this logic: In our approach it is not
some immaterial and self-sufficient geometry (``manifold'') that is presumed to constrain 
the dynamics, but it is the (linear algebra of) dynamics that generates and constrains the 
possible geometry of space-time and determines the form and character of the fields 
(i.e. the bi-vectors) that enable to generate symplectic (``structure-preserving'') 
transformations. 

We have shown that the underlying dynamical system has a representation
by spinors in some abstract phase space which is algebraically separate from the space
of observables: The physical space of observables and generators is related to the dynamical
system like second moments of a distribution are related to the underlying space of random 
variables: The relation is as much of a connection and as it is a separation: If we have
means to change ${\bf F}$ in Eq.~\ref{eq_eqom}, then we change the dynamics of $\psi$, but
what we can observe is not the change of $\psi$, but only the change of ${\bf S}$ 
(Eq.~\ref{eq_envelope}). This is the reason why the conventional description of the
LTs exclusively relates observable quantities. The true nature of the Lorentz transformations
as similarity transformations is uncovered in the context of the Dirac equation only.

\subsection{The Order of Generators}

As listed in Tab.~\ref{tab_order} there are observables of odd (vector, $3$-vectors) and
even (scalar, bi-vectors and pseudo-scalar) order. The multiplications of an arbitrary number
of elements $\y_x$ of even order can only yield elements of even order,
while products involving odd elements can yield all kind of elements. 
Hence the vector elements can be used to produce bi-vectors but not vice versa. We translate
this algebraic fact into a physical interpretation: Matter fields (vectors) can generate 
electromagnetic fields (bi-vectors), but the reverse is impossible: There is no way in this 
formalism to generate matter fields (vectors) using exclusively pure bi-vectors. 
But also a single vector $({\cal E},{\bf p})$ can not be used to generate a bi-vector field, 
since it squares simply to a scalar: Two substantially different vectors are required to 
generate a real bi-vector.

The presented approach is based on the general linear Hamiltonian theory in a Clifford-algebraic
formulation~\cite{qed_paper} and follows a simple and straight logic.
Any Hamiltonian function which is quadratic in the dynamical variables $\psi$ contains a
real symmetric square matrix ${\bf A}$.
The solution of the Hamiltonian equations of motion is based on a real squared skew-symmetric
matrix $\y_0$, called symplectic unit matrix (SUM), which in direct consequence generates
the rules of the algebra Eq.~\ref{eq_cosy_algebra}. They hold for any system of real 
$2\,n\times 2\,n$ (skew-)Hamiltonian matrices. The basic element of phase space in an
abstract degree of freedom. The Dirac algebra is fundamental in the sense that it describes
the simplest general linear kind of interaction, namely linear interaction between two 
degree of freedom.

For a free particle ($\vec E=\vec B=0$), the equations of motion are
\begeq
{d\psi\over d\tau}=({\cal E}\,\y_0+{\vec p}\cdot\vec\y)\,\psi={\bf P}\,\psi\,.
\endeq
and hence
\begeq
{d^2\psi\over d\tau^2}=-({\cal E}^2-{\vec p}^2)\,\psi=-m^2\,\psi\,.
\endeq
which are equivalent to the Dirac and Klein-Gordon equation of a free particle, formulated in 
proper time~\cite{qed_paper,osc_paper} or in other words {\it in the co-moving frame}. 
This becomes more obvious, if we consider the eigenvalues $i\,\omega_\pm$ of ${\bf P}$, 
which are (see App.~\ref{app_eigen}).
\begeq
\omega_\pm=\pm\,\sqrt{{\cal E}^2-\vec p^2}=\pm\,m\,.
\endeq
Hence this basic Hamiltonian theory not only implies the correct
form of the Lorentz transformations of both, space-time coordinates and
electromagnetic fields (including the Lorentz force), it implies the 
relativistic wave equations of QED.

\section{The relativistic Pointing Vector}

The advantage of the Hamiltonian approach becomes apparent,
if the problem is more complicated, for instance in the derivation
of the correct transformation properties of the Pointing vector.
The Pointing (four-) vector represents energy and momentum (-density)
of the electromagnetic field. It is expressed by second order terms
of those fields. In our approach, a central issue of the Pointing
vector is immediately obvious: the square of a Hamiltonian 
bi-vector can only generate scalars, bi-vectors and four-vectors, 
but not vectors.
Hence the Pointing-vector can not be equal to ${\bf F}^2/2$. The
simplest way to construct a Hamiltonian expression of second order
is given by the product of the symmetric second-order matrix
${\bf F}\,{\bf F}^T$ with $\y_0$ as in Eq.~\ref{eq_Smtx}.
With ${\bf F}^T=\y_0\,{\bf F}\,\y_0$ one obtains:
\begeq
  {\bf P}_{e.m.}=({\bf F}\,\y_0\,{\bf F}\,\y_0)\,\y_0/(8\,\pi)
  \label{eq_emE_rest}  
\endeq
With ${\bf F}$ as defined in Eq.~\ref{eq_LorentzForce0} this gives,
written in components, the well-known expressions:
\myarray{
  {\cal E}_{e.m.}&=(\vec E^2+\vec B^2)/(8\,\pi)\\
  {\vec P}_{e.m.}&=\vec E\times \vec B/(4\,\pi)\\
}
But defined this way, ${\bf P}_{e.m.}$ is not Lorentz-covariant {\it unless}
one interprets $\y_0$ here as the four-velocity in the rest-frame
\begeq
{\bf V}=\y\,\y_0+\y\,(\beta_x\,\y_1+\beta_y\,\y_2+\beta_z\,\y_3)\,,
\endeq
such that with $\beta\to 0$ and $\y\to 1$ one has ${\bf V}\to\y_0$.
Then the Lorentz-covariant form should be 
\begeq
  {\bf P}_{e.m.}=-{\bf F}\,{\bf V}\,{\bf F}/(8\,\pi)\,.
  \label{eq_emE_rel}  
\endeq
Written in components Eq.~\ref{eq_emE_rel} gives
\myarray{
  {\cal E}_{e.m.}&={\y\over 4\,\pi}\,\left((\vec E^2+\vec B^2)/2-\vec\beta\cdot(\vec E\times\vec B)\right)\\
  {\vec P}_{e.m.}&={\y\over 4\,\pi}\,\left((\vec E\times \vec B)-(\vec E^2+\vec B^2)\,\vec\beta/2+\right.\\
  &+\left.((\vec\beta\cdot\vec E)\,\vec E+(\vec\beta\cdot\vec B)\,\vec B)\right)\\
  \label{eq_rohrlich}
}
which is identical to the expressions derived by Rohrlich (Eq. 3.23 and Eq. 3.24 in Ref.~\cite{Rohrlich1970}).
However, Eq.~\ref{eq_emE_rel} is considerably simpler and shorter than Eq.~\ref{eq_rohrlich} and
the derivation is, within the suggested approach, straightforward.

\section{Summary}

The presented Hamilton-Clifford-Dirac formalism allows to compute 
the Lorentz-transformation (rotations and boosts) for the ten core
quantities of (charged) particle dynamics (${\cal E},\vec p,\vec E,\vec B$)
simultaneously by symplectic similarity transformations of real 
$4\times 4$-matrices.
Neither does this formalism require complex numbers nor does it require
the use of ``co-'' and ``contra-variant'' vectors or the lifting or
lowering of indices, respectively. Similarity transformations are not
only simpler, they are in a sense more ``natural''.

The suggested matrix formalism for the description of space-time
coordinates and Lorentz transformations provides not only the simplest 
possible and most elegant form of the Lorentz transformations for
the basic physical quantities, i.e. 4-vectors and the six
electromagnetic field components, but also a form that has both,
mathematical and physical significance.
The specific use of real Clifford algebras builds a bridge between 
classical (symplectic) Hamiltonian theory and quantum mechanics. 
It is - as we believe - specifically of high
educational value as it introduces and explains a variety of concepts like 
symplectic motion, linear Hamiltonian systems, group theory, canonical 
transformations, eigenvalues and -vectors, phase space, Lorentz 
transformations, Lorentz force, the Pointing Vector, Clifford algebras, 
the Dirac equation and matrix exponentials by the analysis of the 
algebraic properties of little more than real $4\times 4$-matrices.
Furthermore this approach might be of interest for the use in numerical 
modeling - not because it is faster (it might be, but we did not check),
but mostly because it is simple, stable and ideally suited for modular
programming.

Algebraic equations appear in almost every branch of physics, but cases 
in which a theoretical framework demonstrates the physical  
significance of {\it all} mathematically possible terms are rare. 
In the majority of cases known to the author, the number of algebraically 
{\it possible} terms exceeds the number of physically {\it relevant} terms by far. 
This is different in the presented formalism: There are ten mathematically 
{\it possible} parameters that determine the form of a real Hamiltonian 
$4\times 4$-matrix (Eq.~\ref{eq_eqom}) and all ten parameters have their
specific physical significance.

This type of algebraic integrity provides the proof of maximal simplicity
and the legitimizes to speak of the simplest possible form of the
Lorentz transformations. Futhermore the presented approach allows for
an exceptionally elegant and versatile treatment.

As we have shown, the second moments of a phase space distribution 
of two coupled classical oscillators provide the signature of space-time 
geometry. Elsewhere we argued in some detail that and why the case of the 
real $4\times 4$ matrices is of special significance. Taken serious, this
approach can be argued to provide strong arguments for the apparent 
dimensionality of space-time~\cite{qed_paper,osc_paper}. 

\begin{appendix}

\section{Eigenvalues}
\label{app_eigen}

Using Eq.~\ref{eq_EMEQ}, the eigenvalues $\pm\,i\,\omega_i$ of ${\bf F}$ can
be written as:
\begary{rcl}
K_1&=&{\cal E}^2+\vec B^2-\vec E^2-\vec P^2\\
K_2&=&({\cal E}\,\vec B+\vec E\times\vec P)^2-(\vec E\cdot\vec B)^2-(\vec P\cdot\vec B)^2\\
\omega_1 &=&\sqrt{K_1+2\,\sqrt{K_2}}\\
\omega_2 &=&\sqrt{K_1-2\,\sqrt{K_2}}\\
\label{eq_eigenfreq}
\endary
There are two special cases, the first is a ``inertial status'' (no
accelerations $\vec E=0$ and no rotations $\vec B=0$):
\begary{rcl}
K_1&=&{\cal E}^2-\vec P^2\\
K_2&=&0\\
\omega_1 &=&\omega_2=\sqrt{K_1}\,.
\endary
The other special case is the absence of matter (${\cal E}=0$, $\vec P=0$)
such that
\begary{rcl}
K_1&=&\vec B^2-\vec E^2\\
K_2&=&-(\vec E\cdot\vec B)^2\,,
\endary
which are the known Lorentz invariants of the electromagnetic field.
The (eigen-) frequencies vanish for the standard approach of 
electromagnetic waves (in which $K_1=K_2=0$), which can be
interpreted in such a way that pure electromagnetic waves do not 
constitute a reference frame (i.e. they have no {\it eigenfrequency}).

\end{appendix}

\begin{acknowledgments}
Mathematica\textsuperscript{\textregistered} has been used for parts 
of the symbolic calculations. 
\end{acknowledgments}

\section*{References}

\bibliography{lt_paper_v3}{}

\begin{thebibliography}{10}

\bibitem{LevyLeblond}
Jean-Marc Levy-Leblond.
\newblock One more derivation of the lorentz transformation.
\newblock {\em Am. J. Phys.}, 44(3):271--276, 1976.

\bibitem{Dirac80}
P.A.M. Dirac.
\newblock Why we believe in the einstein theory.
\newblock In B.~Gruber and R.S. Millman, editors, {\em Symmetries in Science}.
  Plenum Press (New York \& London), 1980.

\bibitem{Crowe}
Michael~J. Crowe.
\newblock {\em A History of Vector Analysis}.
\newblock 1994.

\bibitem{Chappell}
James~M. Chappell, Azhar Iqbal, John~G. Hartnett, and Derek Abott.
\newblock The vector algebra war: a historical perspective.
\newblock {\em IEEE Access}, 4:1997--2004, 2016.

\bibitem{Kim1981}
Y.S. Kim and Marilyn~E. Noz.
\newblock Symplectic formulation of relativistic quantum mechanics.
\newblock {\em J. Math. Phys.}, 22(10):2289--2293, 1981.

\bibitem{STA}
D.~Hestenes.
\newblock {\em Space-Time Algebra}.
\newblock Gordon \& Breach, 1966.

\bibitem{GA}
D.~Hestenes and Garret Sobczyk.
\newblock {\em Clifford Algebra to Geometric Calculus}.
\newblock D. Reidel Publ. Company, 1984.

\bibitem{Baylis}
W.E. Baylis.
\newblock Special relativity with $2\times 2$-matrices.
\newblock {\em Am. J. Phys.}, 48:918--925, 1980.

\bibitem{Salingaros}
N.~Salingaros.
\newblock Relativistic motion of a charged particle, the lorentz group, and the
  thomas precession.
\newblock {\em J. Math. Phys.}, 25:706--716, 1984.

\bibitem{GLT}
Sheldon Goldstein, Joel~L. Lebowitz, Roderich Tumulka, and Nino Zanghi.
\newblock On the distribution of the wave function for systems in thermal
  equilibrium.
\newblock {\em J. of Stat. Phys.}, 125(5--6):1197--1225, 2006.

\bibitem{qed_paper}
C.~Baumgarten.
\newblock Relativity and (quantum-) electrodynamics from (onto-) logic of time.
\newblock In Ruth~E. Kastner, Jasmina Jeknic-Dugic, and George Jaroszkiewicz,
  editors, {\em Quantum Structural Studies}. World Scientific, 2017.

\bibitem{osc_paper}
C.~Baumgarten.
\newblock Old game, new rules: Rethinking the form of physics.
\newblock {\em Symmetry}, 8(5), 2016.

\bibitem{ColemanQFT}
Sidney Coleman.
\newblock {\em {Quantum Field Theory, Lectures of}}.
\newblock World Scientific, 2019.

\bibitem{Greiner}
Walter Greiner.
\newblock {\em Classical Mechanics}.
\newblock 2010.

\bibitem{Jackson}
John~David Jackson.
\newblock Classical electrodynamics.
\newblock 1962.

\bibitem{Quaternions}
W.R. Hamilton.
\newblock On quaternions; or on a new system of imaginaries in algebra.
\newblock {\em The London, Edinburgh, and Dublin Philosophical Magazine and
  Journal of Science}, 25(163):10--13, 1844.

\bibitem{exp_paper}
C.~Baumgarten:.
\newblock Analytic expressions for exponentials of specific hamiltonian
  matrices.
\newblock {\em ArXiv:1703.02893}, 2017.

\bibitem{Bott}
Raoul Bott.
\newblock The periodicity theorem for the classical groups and some of its
  applications.
\newblock {\em Adv. In Math.}, 4:353--411, 1970.

\bibitem{Okubo}
Susumu Okubo.
\newblock Real representations of finite clifford algebras.
\newblock {\em J. Math. Phys.}, 32:1657--1674, 1991.

\bibitem{rdm_paper}
C.~Baumgarten.
\newblock Use of real dirac matrices in 2-dimensional coupled linear optics.
\newblock {\em Phys. Rev. ST Accel. Beams}, 14:114002, 2011.

\bibitem{geo_paper}
C.~Baumgarten.
\newblock Geometrical method of decoupling.
\newblock {\em Phys. Rev. ST Accel. Beams.}, 15:124001, 2012.

\bibitem{Goldstein}
H.~Goldstein, C.~Poole, and J.~Safko.
\newblock {\em Classical Mechanics}.
\newblock Addison Wesley, 2002.

\bibitem{Messiah}
Albert Messiah.
\newblock {\em Quantum Mechanics Vol. II}.
\newblock North-Holland Publ., 1961.

\bibitem{Lounesto}
P.~Lounesto.
\newblock Clifford algebras and spinors.
\newblock {\em Cambridge University Press}, 1997.

\bibitem{jacobi}
C.~Baumgarten.
\newblock {A Jacobi Algorithm in Phase Space: Diagonalizing (skew-) Hamiltonian
  and Symplectic Matrices with Dirac-Majorana Matrices}.
\newblock {\em ArXiv:2008.13409v2}, 2021.

\bibitem{Dirac}
P.A.M. Dirac.
\newblock A remarkable representation of the 3+2 de sitter group.
\newblock {\em J. Math. Phys.}, 4(7):901--909, 1963.

\bibitem{Strocchi}
F.~Strocchi.
\newblock Complex coordinates and quantum mechanics.
\newblock {\em Rev. Mod. Phys.}, 38(1):36--40, 1966.

\bibitem{GV}
Stephen~K. Gray and John~M. Verosky.
\newblock Classical hamiltonian structures in wave packet dynamics.
\newblock {\em The Journal of Chemical Physics}, 100:5011--5022, 1994.

\bibitem{Ralston}
John~P. Ralston.
\newblock $\cal{PT}$ and $\cal{CPT}$ quantum mechanics embedded in symplectic
  quantum mechanics.
\newblock {\em J. Phys. A: Math. Theor.}, 40:9883--9904, 2007.

\bibitem{Khrennikov}
Andrei Khrennikov.
\newblock Quantum mechanics from statistical mechanics of classical fields.
\newblock {\em J. Phys.: Conf. Ser.}, 70(2007):012009, 2007.

\bibitem{Briggs}
John~S. Briggs and Alexander Eisfeld.
\newblock Coherent quantum states from classical oscillator amplitudes.
\newblock {\em Phys. Rev. A}, 85.

\bibitem{Klyshko}
D.N. Klyshko.
\newblock The bell theorem and the problem of moments.
\newblock {\em Phys. Lett. A}, 218:119--127, 1996.

\bibitem{fit_paper}
C.~Baumgarten.
\newblock The final theory of physics - a tautology?
\newblock {\em ArXiv:abs/1702.00301}, 2017.

\bibitem{Schmueser}
P.~Schm\"user.
\newblock {\em Feynman-Graphen und Eichtheorien f\"ur Experimentalphysiker},
  volume 295 of {\em Lecture Notes in Physics}.
\newblock Springer, 1988.

\bibitem{Hestenes}
D.~Hestenes.
\newblock Mysteries and insights of dirac theory.
\newblock {\em Annales de la Fondation Louis de Broglie}, 28(3-4):367--389,
  2003.

\bibitem{MHO}
K.R. Meyer and D.~Offin.
\newblock {\em Introduction to Hamiltonian Dynamical Systems and the N-Body
  Problem}.
\newblock Springer, New York, 2017.

\bibitem{Kim}
Y.S. Kim and M.E. Noz.
\newblock Dirac matrices and feynman's rest of the universe.
\newblock {\em Symmetry}, 4:626--643, 2012.

\bibitem{Feynman}
Richard~P. Feynman, Robert~B. Leighton, and Matthew Sands.
\newblock Feynman lectures vol. iii.

\bibitem{Lax}
Peter~D. Lax.
\newblock Integrals of nonlinear equations of evolution and solitary waves.
\newblock {\em Comm. Pure Appl. Math.}, 21(5):467--490, 1968.

\bibitem{Rohrlich1970}
Fritz Rohrlich.
\newblock Electromagnetic momentum, energy and mass.
\newblock {\em Am. J. Phys.}, 38(11):1310--1316, 1970.

\end{thebibliography}
\bibliographystyle{unsrt}

\end{document}